\documentclass[twocolumn]{aastex701}
\shorttitle{Cool Explosions}
\shortauthors{A.Antoni, Y.-F.Jiang, and E.Quataert}

\usepackage{CJK}
\usepackage{xcolor}
\usepackage{booktabs} 
\usepackage{amsmath}	
\newcommand{\beq}{\begin{equation}}
\newcommand{\eeq}{\end{equation}}
\newcommand{\beqn}{\begin{equation*}}
\newcommand{\eeqn}{\end{equation*}}
\newcommand{\ath}{\texttt{Athena++} }
\newcommand{\mesa}{\texttt{MESA} }
\newcommand{\athns}{\texttt{Athena++}}
\newcommand{\mesans}{\texttt{MESA}}
\newcommand{\msun}{M_\odot}
\newcommand{\rsun}{R_\odot}

\newcommand{\rin}{r_{\rm in}}
\newcommand{\Mbh}{M_{\rm BH}}
\newcommand{\edep}{E_{\rm dep}}
\newcommand{\extra}{E_{\rm xtra}}
\newcommand{\ebind}{E_{\rm bind}}
\newcommand{\lsun}{L_\odot}
\newcommand{\mzams}{M_{\rm zams}}
\newcommand{\menv}{M_{\rm env}}
\newcommand{\mconv}{M_{\rm conv}}
\newcommand{\rexp}{r_{\rm exp}}
\newcommand{\hexp}{h_{\rm exp}}
\newcommand{\mej}{M_{\rm ej}}
\newcommand{\eej}{E_{\rm ej}}
\newcommand{\vej}{v_{\rm ej}}
\newcommand{\lpl}{L_{\rm pl}}
\newcommand{\tpl}{t_{\rm pl}}
\newcommand{\teff}{T_{\rm eff}}

\definecolor{andrea}{HTML}{FF376A}
\definecolor{myblue}{HTML}{1E90FF}

\begin{document}

\title[Cool explosions]{Radiation Hydrodynamic Simulations of Low-Energy Explosions of Red and Yellow Supergiants}

\author[orcid=0000-0003-3062-4773,gname='Andrea',sname='Antoni']{Andrea Antoni}
\affiliation{Center for Computational Astrophysics, Flatiron Institute, New York, NY 10010, USA}
\email[show]{aantoni@flatironinstitute.org}  

\author[orcid=0000-0002-2624-3399,gname='Yan-Fei',sname='Jiang']{Yan-Fei Jiang (\begin{CJK*}{UTF8}{gbsn}姜燕飞\end{CJK*})} 
\affiliation{Center for Computational Astrophysics, Flatiron Institute, New York, NY 10010, USA}
\email{yjiang@flatironinstitute.org}

\author[orcid=0000-0001-9185-5044,gname=Eliot,sname=Quataert]{Eliot Quataert}
\affiliation{Department of Astrophysical Sciences, Princeton University, Princeton, NJ 08544, USA}
\email{quataert@princeton.edu}

\begin{abstract}
A variety of physical processes leads to the low-energy ejection of material from the hydrogen-rich envelopes of red and yellow supergiants.  These include common envelope events, stellar mergers, eruptive mass loss, and failed supernovae. These events may appear as luminous red novae, intermediate luminosity red transients, supernova imposters, or other transients with similar lightcurves and colors that are followed by the disappearance of the progenitor star (e.g. failed supernovae). The Vera C. Rubin Observatory will find these events in large numbers; detailed modeling of their lightcurves is essential for photometrically differentiating between these important physical processes in the lives of massive stars.  We use one-dimensional, radiation hydrodynamic simulations to model the lightcurves of low-energy explosions of red and yellow supergiants. Red supergiant explosions have durations of 100-400 days, longer than Type IIp supernovae, while stripped, yellow supergiant explosions have durations of 10s of days.  Our models probe the boundary between the radiation-pressure dominated and gas-pressure dominated regimes.  We provide fitting formulae for the plateau luminosity and duration of the events. Finally, we show that the failed supernovae candidates in NGC 6946 and M31 are consistent with failed supernovae models for explosion energies of $\sim10^{47}-10^{49}$ erg.

\end{abstract}

\keywords{\uat{Stellar mass black holes}{1611} ---
\uat{Massive stars}{732} ---
\uat{Supernovae}{1668} ---
\uat{Core-collapse supernovae}{304}}


\section{Introduction}

There are a variety of intermediate-luminosity transients of massive-star origin that involve the ejection of hydrogen-rich material with a wide range of energies spanning the ``gap'' between supernovae (SN) with peak luminosities $\gtrsim 10^{42}$ erg/s and classical novae with luminosities $\lesssim 10^{39}$ erg/s \citep{2011PhDT........35K, 2022Univ....8..493C}.  Among these are the brighter population ($-$9 to $-$15 mag) of extra-Galactic luminous red novae \citep[LRNe, ][]{2023ApJ...948..137K}, intermediate luminosity red transients \citep[ILRTs, ][]{2021A&A...654A.157C},  pre-SN outbursts \citep[the SN ``imposters,''][]{2002PASP..114..700V}, low-luminosity Type IIp SNe \citep{2025PASP..137d4203D,2025arXiv250620068D}, and failed SNe \citep{2017MNRAS.468.4968A,2024arXiv241014778D}. 

LRNe are generally thought to be associated with stellar mergers or common-envelope events in which orbital energy is used to eject a substantial fraction of the donor star's envelope. Brighter events are likely associated with more massive donors. Indeed, several extra-Galactic LRNe have been observed to have yellow supergiant (YSG) or blue supergiant progenitors \citep{2016MNRAS.458..950S,2017ApJ...834..107B,2019A&A...630A..75P,2021A&A...653A.134B}.  ILRTs are most often attributed to terminal, electron-capture SN of super asymptotic giant branch stars \citep{2021A&A...654A.157C,hiramatsu_electron-capture_2021}. Pre-SN outbursts (broadly encompassing non-terminal eruptive mass loss from an evolved massive star) can appear as a distinct class, though some LRNe or ILRTs may be eruptive mass-loss events \citep[e.g. NGC 4490-OT2011 and AT2019krl, ][]{2016MNRAS.458..950S,2021ApJ...917...63A}.  

In ``failed'' SNe, the collapse of a massive star's core does not lead to a canonical neutrino-driven, turbulence-aided, $\sim10^{51}$ erg explosion of the star\footnote{In the jittering jets explosion mechanism framework, \cite{2024Univ...10..458S} argues there are no failed supernovae.}. Rather, the inner parts of the star collapse to form a black hole (BH) and a large fraction of the star remains bound to the BH \citep[e.g., ][]{2011ApJ...730...70O,burrows_channels_2025}. Just prior to BH formation, neutrino cooling reduces the effective gravitational mass of the hot PNS star.  This nearly instantaneous change in the gravitational potential results in the envelope being over-pressured; a sound pulse forms that steepens into a weak shock in the outer parts of the star \citep{1980Ap&SS..69..115N,Lovegrove2013,2018MNRAS.477.1225C}. This weak shock can unbind $\sim$few $\msun$ of the outer RSG envelope and $\sim10^{-2}\msun$ of the YSG envelope \citep[][the amount of mass lost is sensitive to the assumed equation of state of the PNS and the structure of the star at collapse]{2018MNRAS.476.2366F,2021ApJ...911....6I, 2023ApJ...942...16D}.  Meanwhile, if the core is not rapidly rotating, as is probably the case for RSGs and YSGs, the inner parts of the star (including the helium layer) falls in and accretes onto the newly-formed BH \citep{paperI}.  Any of the convective hydrogen layer that is still bound will begin to fall in, but will face a significant barrier to accretion due to the angular momentum associated with the random velocity field in the convective layer \citep{paperI}.  Roughly $~1\%$ of the convective hydrogen layer can accrete. The rest flows back out and the small amount of accretion allows conversion of the potential energy at the circularization radius into a $~10^{47}$ to $~10^{49}$ erg explosion of the rest of the envelope gas \citep{paperII}.  Observationally, the ejection of the envelope may give rise to an optical outburst or a brightening in the infrared as the low-energy ejecta quickly cools and forms dust.  

Although these bright, hydrogen-rich transients arise through diverse channels, the outbursts themselves are driven by a common mechanism: the injection of $\sim$10$^{46}$ to $\sim$10$^{50}$ erg of energy over a finite timescale into the cool, hydrogen envelope of an evolved massive star.  If the energy injection is rapid, this leads to a weak shock that propagates outward and may unbind all or part of the envelope.  As the ejecta expand and cool, the photosphere is set by hydrogen recombination, as in a Type IIP SN. The event will show a IIP-like plateau, though lasting much longer because the expansion velocities are only $100$s to $\sim 1000$ km/s (as compared to $\sim$4,000 km/s for IIP ejecta).  If the energy is injected over longer timescales, it may drive an outburst with very little mass (and could be more like a wind than an eruption).  Because similar physics sets the observable properties of these transients, there is a critical need for detailed models to distinguish, observationally, between these ways of removing mass from the stellar envelope. 

As an initial effort, we model the simplest scenario, which is the instantaneous explosion of the hydrostatic envelopes of RSGs and partially-stripped YSGs with energies from $10^{47}$ to $10^{50}$ erg. Previous radiation-hydrodynamical studies of lower-energy explosions have focused on RSG progenitors (\citealt{2010MNRAS.405.2113D,Lovegrove2017,tsuna_transients_2025}; though also see the semi-analytic lightcurve models of \citealt{2022ApJ...938....5M} which models LRNe ejecta independently from the progenitor).  Motivated by the partially-stripped YSG progenitor of the failed SN candidate in M31 \citep{2024arXiv241014778D}, the preponderance of YSG progenitors of LRNe, and the likely partially-stripped progenitors of Type IIb SNe, we include a set of partially-stripped stars that die as YSGs in our study.

Modeling hydrogen-rich explosions at SN energies ($\sim$10$^{51}$ erg) permits some simplifications. Namely, one can assume the gas is radiation-pressure-dominated behind the shock, the energy deposited by the shock is orders-of-magnitude larger than the other terms in the energy equation, and the opacity behind the shock and behind the photosphere on the plateau is dominated by electron scattering.  At the energies of interest here, these assumptions break down.  Instead, the gravitational binding energy of the envelope is dynamically important as the shock climbs out of the envelope \citep{2018MNRAS.477.1225C} and the absorption opacity can be similar to or larger than the electron scattering opacity on the plateau \citep{Lovegrove2017}. In addition,  radiation pressure is not as dominant as in higher-energy explosions \citep{2019ApJ...879...20F}; indeed we will find that our models straddle the radiation- and gas-pressure-dominated regimes.  To capture this more complex physical picture, we perform fully time-dependent  radiation hydrodynamic (RHD) simulations with realistic stellar opacities using \athns. \ath solves the full RHD equations for the specific intensities over discrete ordinates providing accurate solutions in both optically thick and optically thin conditions. 

We note that a major goal of this work is to develop the methodology to study a variety of problems involving outbursts or explosions with significant fallback accretion in \athns. Eulerian codes are better suited for handling accretion flows  than Lagrangian codes as material can leave the domain without requiring a remapping of the grid. An additional benefit to developing this methodology in \ath is to be able to employ the code's sophisticated implementation of radiation hydrodynamics and, in the future, the recently-developed multi-group capabilities of \ath \citep{Jiang2022}.  The present work studying instantaneous explosions of supergiant envelopes is an initial application of this methodology; we will apply it to other problems in the future.

This paper is organized as follows.  Sec.~\ref{sec:mesamodels} describes our method for generating model RSGs and YSGs with \mesans.  Sec.~\ref{sec:athenamethods} details our method for simulating weak explosions of the stars with \athns.  We present our results in Sec.~\ref{sec:results}. Sec.~\ref{sec:discussion} compares our results to previous work and discusses the limitation of our models. Finally, Sec.~\ref{sec:summary} summarizes our main results and places our models in the context of upcoming time-domain surveys.

\section{Progenitor models}
\label{sec:mesamodels}
We use version 24.08.1 of the Lagrangian stellar evolution code \mesa \citep{2011ApJS..192....3P,2013ApJS..208....4P,2015ApJS..220...15P,2018ApJS..234...34P,2019ApJS..243...10P,2023ApJS..265...15J} to generate a suite of massive star envelopes at the end of core carbon burning. We generate two sets of non-rotating stellar models. Our H-rich models are RSGs that have retained most of their hydrogen envelopes. Our H-poor models are YSGs that have lost all but $<1.0$ $\msun$ of their hydrogen envelopes.  We stop the models at the end of core carbon burning because the envelope does not change substantially during the rest of the evolution of the star beyond carbon burning.  Thus these models represent the stars at the time of core collapse. 

We use a setup similar to the test suite problem \texttt{20M\_pre\_ms\_to\_core\_collapse} supplied with version 24.08.1 of \mesa but with the following changes.  We adopt Opal opacities (\texttt{kap\_file\_prefix} $=$ \texttt{oplib\_agss09}) to match the opacity tables of our \ath setup. We adopt a mild wind (\texttt{Dutch\_scaling\_factor} = 0.2), which results in mass-loss rates between $\sim 10^{-9} - 10^{-6} \msun/$yr. Most of the mass-loss happens near the end of core helium burning. It is useful to include part of the atmosphere of the star beyond an optical depth of 1 when mapping the envelope from \mesa to \athns. To achieve this, we allow the model to relax to an optical depth of $10^{-2}$. Specifically, we include \texttt{relax\_tau\_factor\_after\_core\_c\_burn} $=5 \times 10^{-4}$, \texttt{relax\_to\_this\_tau\_factor} $=0.01$, and \texttt{dlogtau\_factor} $= 0.1$ in \texttt{inlist\_to\_end\_core\_c\_burn} and allow the model to run until the central carbon-12 mass fraction is $10^{-5}$. We calculate models for zero-age main sequence (ZAMS) masses between $12\msun$ and $20\msun$ in increments of $2\msun$.

The H-poor and H-rich models are identical until the end of core He burning.  For the H-poor models, we use the provided \texttt{inlist\_remove\_envelope} to instantaneously strip the hydrogen envelope after core He burning down to a hydrogen envelope mass of \texttt{extra\_mass\_retained\_by\_remove\_H\_env}.  After stripping the envelope, we continue the model to the end of core C burning identically to the H-rich models.  We note that, in nature, post-red giant YSGs can occupy a large area in the HR diagram depending on how the star is losing mass and how long ago the mass was lost.  If one wishes to study YSGs that are in the process of losing their envelope (either due to stripping by a binary companion as the primary expands up the red giant branch or a strong RSG wind), then the structure, color, and luminosity of the YSG envelope would depend both on the rate of mass-loss and the mass-loss history (i.e. how long ago the mass-loss episode happened).  In our case, we are interested in the structure of the star at the end of its life, which is decoupled from the particular scenario that caused the star to end its life with a low mass envelope (thus, dying as a YSG instead of a RSG). We find that, by the end of core C burning, models stripped via a wind during core He burning end up with the same $L$, $T_{\rm eff}$, He core mass, and envelope structure as our models stripped instantaneously after core He burning.  Although the path the star takes to that state is different, the final state is the same. Thus, we adopt the simpler method of instantaneous removal for this study.  

\begin{figure}
\centering
\includegraphics[width=0.9\columnwidth]{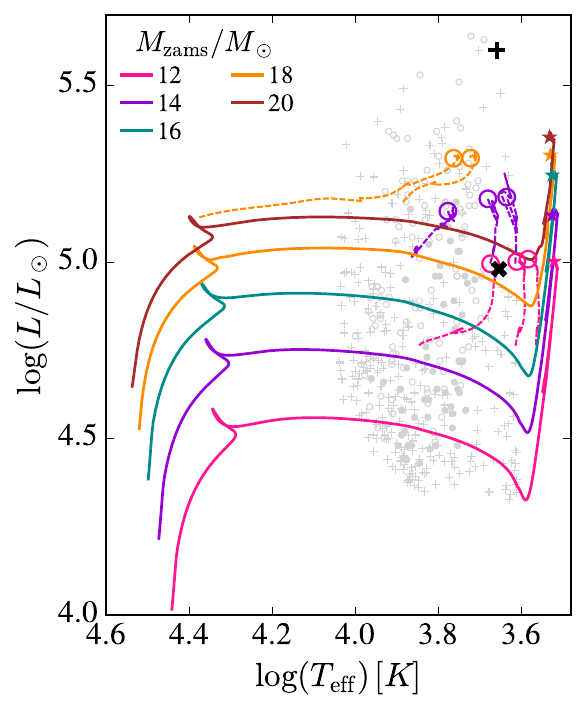}
\caption{Evolutionary tracks of our \mesa models through the end of core C burning. Color indicates the zero-age-main sequence mass of the star. Solid curves show tracks for the unstripped stars. Dashed lines show the stripped stars post-stripping. Filled stars (open circles) mark the final state of the stripped (unstripped) models.  The stripped models end their lives at hotter temperatures the lower their hydrogen envelope mass. 
 The bold x and bold plus, respectively, mark the locations of the progenitors of the failed supernova candidates in M31 and NGC 6946 \citep{2024arXiv241014778D}. For comparison, the grey points in the background show the populations of YSGs in M31 \citep[filled circles, ][]{drout_yellow_2009}, M33 \citep[open circles, ][]{drout_yellow_2012}, and the Large Magellanic Cloud \citep[pluses, ][]{2012ApJ...749..177N}. }
\label{fig:mesahr}
\end{figure}
We summarize our suite of progenitor models in Table \ref{tab:mesamodels}, providing important parameters of the model at the end of core C burning, which is the state of the model used to initialize our \ath simulations.  Not all of these models are utilized in our \ath simulations, but we include the final parameters for completeness and to show how these parameters depend on ZAMS mass and envelope mass. Figure 1 shows evolutionary tracks of the models. Color indicates the ZAMS mass of the model, as noted in the legend. The solid curves show the tracks for the unstripped stars up to the end of core C burning (filled stars), which is the point at which we hand off the model to \athns. These models die as RSGs.  The dashed curves show the evolution of the stripped stars after we instantaneously reduce the mass of the envelope (the track up to the end of core He burning is identical to the unstripped models). As the stars recover thermal equilibrium, they move up and to the right. By the end of core C burning (open circles), they return to roughly the same luminosity as the analogous RSG, but with hotter surface temperature.  This is due to the fact that, even though most of the envelope is still convective, the envelope becomes more radiatively efficient as the hydrogen envelope mass is reduced.  As the envelope mass is reduced, therefore, the envelope contracts and the star shifts leftward to hotter temperatures and away from the RSG track.  Each of the dashed tracks of the same color corresponds to a different final envelope mass for the same ZAMS mass, with lower mass envelopes falling at higher temperatures.  Thus, the open circles fall in order of increasing envelope mass from left to right. See Table \ref{tab:mesamodels} for the envelope mass of each of these models for each family of stars with the same ZAMS mass.

\begin{deluxetable*}{lccccccccccc}
\tablecaption{Table of \mesa progenitor models.\label{tab:mesamodels}}
\tablehead{
Model & $\mzams$  & $M_\star$  & $R_\star$ & $L_\star$ & $M_{\rm core}$ & $\menv$ & $M_{\rm conv}$ & $r_{\rm exp}$ &  $|\ebind(r_{\rm exp})|$ & $\mu$ & $X_{\rm H}$ \\
	  Name & $[\msun]$ & $[\msun]$ & $[\rsun]$ & $[10^5\lsun]$ & $[\msun]$ & $[\msun]$  & $[\msun]$ &$[\rsun]$ & [$10^{46}$ erg] \\
}
\startdata 
\multicolumn{3}{l}{\it H-rich (unstripped) models} \\
\hline
12 & 12&    11.5 &     958 &     1.0 &     4.5 &     7.0 &     6.9 &     8.1 &    20.9&    0.64 &    0.67 \\
14 & 14&    13.2 &    1090 &     1.4 &     5.6 &     7.6 &     7.4 &     5.2 &    23.5&    0.64 &    0.66 \\
16 & 16&    14.9 &    1243 &     1.8 &     6.7 &     8.2 &     8.0 &     4.1 &    24.9&    0.65 &    0.65 \\
18 & 18&    16.5 &    1298 &     2.0 &     7.2 &     9.3 &     9.2 &     4.0 &    30.5&    0.66 &    0.63 \\
20 & 20&    18.5 &    1365 &     2.3 &     7.8 &    10.6 &    10.5 &     3.4 &    37.6&    0.66 &    0.63 \\
\hline
\multicolumn{3}{l}{\it H-poor (partially-stripped) models} \\
\hline
12.3 & 12&     4.8 &     470 &     1.0 &     4.5 &     0.3 &     0.2 &    13.8 &     0.7&    0.69 &    0.58 \\
12.5 & 12&     5.0 &     633 &     1.0 &     4.5 &     0.5 &     0.4 &    13.2 &     1.0&    0.66 &    0.63 \\
12.7 & 12&     5.2 &     723 &     1.0 &     4.5 &     0.7 &     0.6 &    12.9 &     1.3&    0.65 &    0.64 \\
14.3 & 14&     5.9 &     347 &     1.4 &     5.6 &     0.3 &     0.1 &    16.5 &     1.1&    0.75 &    0.48 \\
14.5 & 14&     6.1 &     564 &     1.5 &     5.6 &     0.5 &     0.3 &    14.4 &     1.4&    0.68 &    0.59 \\
14.7 & 14&     6.3 &     699 &     1.5 &     5.6 &     0.7 &     0.5 &    14.1 &     1.7&    0.67 &    0.62 \\
18.5 & 18&     7.7 &     440 &     2.0 &     7.2 &     0.5 &     0.4 &    14.6 &     3.0&    0.78 &    0.44 \\
18.7 & 18&     7.9 &     532 &     2.0 &     7.2 &     0.7 &     0.6 &    14.7 &     3.8&    0.77 &    0.44 \\
	  \hline
\enddata
\tablecomments{\\Model name -- For the stripped models, the decimal number denotes $M_{\rm env}$, rounded to the tenths place for readability. \\
	$\mzams$ -- Zero-age-main-sequence mass. \\ 
        The remaining columns refer to the evolved star (at the end of core C burning) used in our \ath models:\\
	$M_\star$ -- Stellar mass. \\
	$R_\star$ -- Stellar radius. \\ 
        $M_{\rm core}$ -- Mass of the He core.\\
	$\menv$ -- Mass of the envelope (as defined by \mesa), which is the mass outside the He core 
        (roughly where $X_{\rm H} > 0.1$). \\
        $M_{\rm conv}$ -- Mass of the inert, convective hydrogen envelope (where $X_{\rm H} > 0.43$) that we explode in our \ath models. This roughly maps to the mass of ejecta as we deposit enough energy to unbind this mass.\\
	$r_{\rm exp}$ -- Radius where we place the thermal bomb, which is at mass coordinate $M_\star - M_{\rm conv}$. \\ 
	$|\ebind(r_{\rm exp})|$ -- Total thermal and gravitational energy of the mass with $r > r_{\rm exp}$.  We deposit $E_{\rm dep} = |\ebind| + E_{\rm xtra}$ in the thermal bomb so that $E_{\rm xtra}$ roughly maps to the asymptotic energy of the ejecta. \\
    $\mu$ -- Mean molecular weight of the bulk of the convective hydrogen envelope. \\
    $X_{\rm H}$ -- Hydrogen mass fraction of the bulk of the convective hydrogen envelope.}
\end{deluxetable*}

\section{Simulation Methods}
\label{sec:athenamethods}
In this work, we simulate low-energy explosions of the hydrogen envelopes of RSGs and YSGs in spherical symmetry.  The initial envelope profiles are mapped into \ath from the model RSGs and YSGs described in Sec.~\ref{sec:mesamodels}. By including realistic stellar opacities and solving the time-dependent radiation hydrodynamic equations, and including the self-gravity of the gas, our methods are well-suited for these low-energy explosions for which all of the terms in the energy equation are dynamically important and are thus challenging to model analytically.  In this section, we describe methods for simulating these explosions in \athns.  

\subsection{Equations solved} 
We use the Eulerian radiation hydrodynamics (RHD) code \athns\footnote{Version 21.0, \href{https://github.com/PrincetonUniversity/athena}{https://github.com/PrincetonUniversity/athena}} \citep{2020ApJS..249....4S} to solve the coupled equations of inviscid hydrodynamics and frequency-integrated radiation transport \citep{Jiang2014,Jiang2021}.  Our one-dimensional simulation domain adopts spherical symmetry and assumes local thermal equilibrium for the gas emission term.  All models employ second-order time integration (Van-Lear, \texttt{integrator = VL2}) and spatial reconstruction (piecewise linear, \texttt{xorder = 2}), the Harten-Lax-van Leer contact Riemann solver (\texttt{$-$$-$flux hllc}), and the implicit solver for the radiative transfer equations, described in \citet{Jiang2021}. \ath solves for the specific intensities over discrete angles. We adopt the spherical-polar angular system described in sec.~3.2.4 of \citealt{Jiang2021}, employing 10 angles for the specific intensity per radial grid zone.   

The (lab frame) RHD equations that we solve are
\beq
\frac{\partial \rho}{\partial t} + \nabla\cdot(\rho {\bf v}) = 0
\eeq

\beq
\frac{\partial (\rho{\mathbf v})}{\partial t} + \nabla\cdot(\rho \mathbf{v}\mathbf{v} + P \textbf{\textsf{I}}) = - \rho \nabla \Phi -{\mathbf S_r}({\mathbf P})
\label{eq:momentum}
\eeq
\beq
\frac{\partial \epsilon}{\partial t} + \nabla\cdot\big[(\epsilon + P)\mathbf{v}\big] = -\rho \mathbf{v}\cdot \nabla \Phi -S_r(\epsilon),
\label{eq:energy}
\eeq

\beq
\frac{\partial I}{\partial t} + c{\mathbf n}\cdot \nabla I =  c S_I.
\label{eq:intensity}
\eeq
In the above equations, $\rho$ is the mass density, $\rho \mathbf{v}$ is the momentum density, $P$ is the gas pressure, \textbf{\textsf{I}} is the identity tensor, $I$ is the lab-frame specific intensity of the radiation, ${\mathbf n}$ is the photon propagation direction, and $c$ is the speed of light.  The total energy density of the gas is $\epsilon = \epsilon_g + \rho \mathbf{v}\cdot\mathbf{v}/2$, where $\epsilon_g$ is the internal energy of the gas.  We adopt the \ath adiabatic equation of state for the gas for which $\epsilon_g = P/(\gamma_g -1)$ with $\gamma_g = 5/3$. The equation of state does not include hydrogen recombination (see discussion in Sec.~\ref{sec:recombination}). The radiation source terms $S_I$, ${\mathbf S_r}({\mathbf P})$, and $S_r(\epsilon)$ are given in eq. (9) of \citet{Jiang2021}. \ath adopts a mixed-frame approach so that the radiation source terms (and thus the opacities) are computed in the co-moving frame, then a frame-transformation is applied to transform the source terms to the lab frame, where they are then added to the lab-frame hydrodynamic equations (i.e., the left-hand sides of eqs. \eqref{eq:momentum} and \eqref{eq:energy}).

\subsection{Opacities}
The radiation source terms depend on the Rosseland mean absorption opacity, $\kappa_a$, the scattering opacity, $\kappa_s$, and the Planck mean opacity, $\kappa_P$.\footnote{Note that $\kappa_P$ is equivalent to $\kappa_{\delta P} + \kappa_a$ in \citet{Jiang2021}.}  Throughout, we assume $\kappa_s$ is due to electron scattering, $\kappa_s = 0.2(1 + X_{\rm H})$, where $X_{\rm H}$ is the hydrogen mass fraction. For temperatures above $\sim 9200$ K, we compute $\kappa_a$ and $\kappa_P$ by interpolating the LANL atomic opacity tables  \citep{2016ApJ...817..116C}, which we compiled using the online TOPS tool.\footnote{\href{https://aphysics2.lanl.gov/apps/}{https://aphysics2.lanl.gov/apps/}} Although the atomic opacities do not depend strongly on $X_{\rm H}$, we nevertheless use two different opacity tables, one with $X_{\rm H} = 0.65$ and one with $X_{\rm H} = 0.5$. A given \ath run adopts the value of $X_{\rm H}$ that is closer to that of the corresponding \mesa model (given in the last column of Table \ref{tab:mesamodels}).  At lower temperatures, we smoothly transition to the low-temperature opacity table of \citet{2021MNRAS.508..453Z}, which is a combination of LANL atomic opacities \citep{2016ApJ...817..116C}, molecular opacities \citep{2014ApJS..214...25F}, and dust opacities that were calculated following \citet{2018ApJ...869L..45B}. \citealt{2021MNRAS.508..453Z} provides tables for three different maximum dust grain sizes assuming Solar abundances; we adopt the smallest maximum grain size available.  We note that the opacities at these temperatures depend very little on $X_{\rm H}$, so the same low-temperature table is used for all of our models. In most cases, the ejecta in our models do not reach dust temperatures ($\lesssim$1000-1500 K, depending on the density) until after the end of the plateau in the light curve, though in our lowest-energy RSG models, dust opacities contribute to the shape of the lightcurve tail. In the mixed-frame approach of \athns, opacities are computed in the co-moving frame and applied in the radiation source terms, which are also computed in this frame. 

\subsection{Self-gravity}
Self-gravity is included via the source term 
\beq
-\nabla\Phi = -GM(r)/r^2
\eeq 
where $M(r) = 4\pi\int_{\rin}^r r^2\rho dr + \Mbh$ is the mass enclosed at $r$ and is updated at each time step.  The mass interior to the inner boundary at $\rin$, $\Mbh$, is evolved as matter is accreted through the absorbing inner boundary at $\rin$. At initialization, all of the mass interior to $\rin$  in the original stellar model is included in $\Mbh$. All models adopt $\rin = 0.1 \rsun$, which falls well interior to the hydrogen envelope in all models.  

\subsection{Boundary Conditions}
Our simulations are performed in spherical symmetry. The inner and outer boundary conditions in the $r$-direction are as follows.

The outer radial boundary condition on the hydrodynamic variables is a zero-gradient, mass-conserving, diode that copies the pressure and positive radial velocities to the ghost zones. If the radial velocity is negative in the last live zone, the radial velocities in the ghost zone are set to zero. The density in the ghost zones is copied from the last live zone but multiplied by a factor of $(r_{\rm live}/r_{\rm ghost})^2$ to conserve mass flux. Outgoing lab-frame specific intensities are copied to the ghost zones; the incoming specific intensities are set to zero in the ghost zones.

In all of our models, the inner boundary is set well interior to the base of the hydrogen envelope, where we place our explosion source. Thus all models have mass at small radii that is bound and will fall in supersonically to small $r$.  Our radial inner boundary conditions facilitate this supersonic flow towards the BH.  Density and velocity are treated similarly as at the outer boundary: we disallow inflow with a diode condition on $v$ but conserve mass by scaling the density by $(r_{\rm live}/r_{\rm ghost})^2$. However, instead of a zero-gradient condition, we floor the gas pressure in the ghost zones to facilitate super-sonic outflow from the domain. At the inner boundary, the gas is very optically thick.  We conserve the outgoing comoving specific intensities, which are then boosted to the lab frame using the boundary conditions on the velocity. 

\subsection{Units}
\label{secr:units}
As described in \citet{Jiang2021}, the radiation module uses dimensionless units based on the dimensionless speed of light, $\mathbb{C} = c/ v_0$, and dimensionless radiation pressure,  $\mathbb{P} = a_r T_0^4 / P_0$, where $T_0$ is the unit of gas temperature, $v_0$ is the speed unit for the gas variables, and $P_0$ is the unit of gas pressure.  Radiation energy density is in units of $a_r T_0^4$ and radiation flux is in units of $ca_r T^4_0$.   At runtime, the user can set $\mathbb{C}$ and $\mathbb{P}$ directly, or compute them from $T_0$, $P_0$, and $v_0$.  For our explosion calculations, we adopt density units of $\rho_0 = 2.233 \times 10^{-8}$ g/cm$^3$, length units of $r_0 = \rsun$, and $T_0 = 1.8028 \times 10^5$ K. We assume an ideal gas, thus we set $P_0 = \rho_0 k_b T_0/ (\mu m_p)$ and $v_0 = (P_0 / \rho_0)^{1/2}$.
Through $P_0$, the balance between gas and radiation pressure at a given temperature thus depends on $\mu$.  Our hydrogen envelopes of our progenitor models have a wide range of $\mu$. In order to initialize the \mesa model in hydrostatic equilibrium, we must choose $\mu$ to match the value from the \mesa model.  When we map a given \mesa progenitor into \athns, we take $\mu$ to be the volume-weighted average $\mu$ over the envelope in the \mesa model (given in the second-to-last column in Table \ref{tab:mesamodels}). As the units thus vary between different models, we will convert to cgs or Solar units when presenting simulation results.

\subsection{Initialization of the \ath simulations}
\begin{figure*}
\centering
\includegraphics[width=0.9\textwidth]{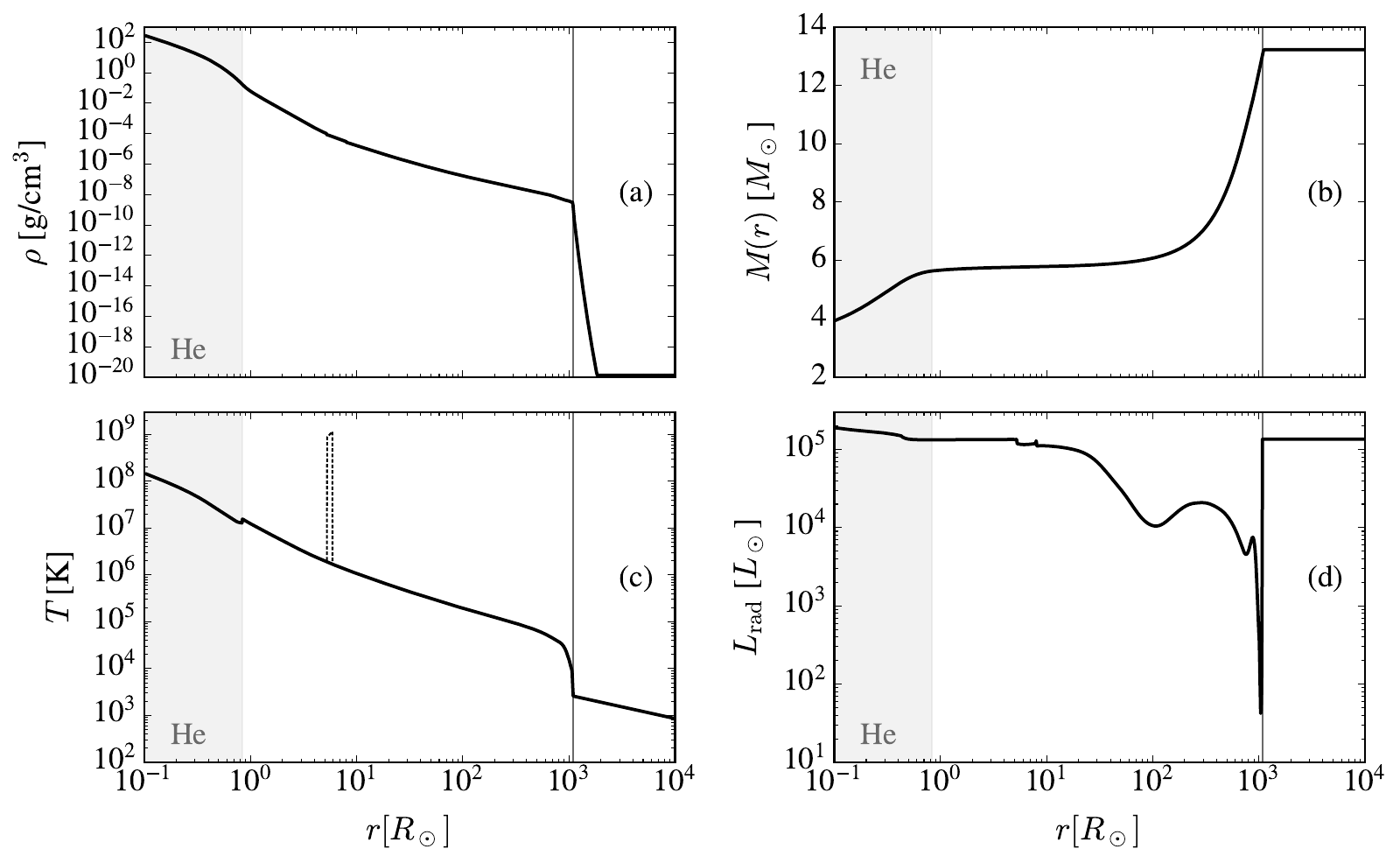}
\caption{Example initialization of gas and radiation in \ath from the $\mzams = 14\msun$ \mesa model. To the left  of the vertical line at $R_\star$, the grid is initialized from the \mesa model. To the right of the vertical line, we initialize an isothermal atmosphere which transitions to a constant density background at $\rho = 10^{-20}$ g/cm$^3$.  The explosion is initialized by adding additional thermal pressure of total energy $\edep$.  The dotted line in the temperature panel shows the extra bump due to a thermal bomb corresponding to a $\sim 10^{48}$ erg explosion.  The thermal bomb is placed at the base of the convective, inert hydrogen envelope, where $X_{\rm H}$ becomes larger than $0.43$ in the \mesa model. The gray shaded region highlights material that was part of the He core in the \mesa model, where $X_{\rm H} < 0.1$.  The He core and most of the material with $X_{\rm H} < 0.43$ falls back and accretes while the pressure bomb matures into a shock and climbs out of the envelope at larger radii.}
\label{fig:initial_profiles}
\end{figure*}

In \athns, we initialize the density and gas pressure using the values from the \mesa profile up to $R_\star$.  At $R_\star$, we switch to an isothermal atmosphere with $T = T_\star$.  The density falls off exponentially in the isothermal atmosphere, but we truncate the fall-off to $\rho = 10^{-20}$ g/cm$^3$. Outside of this radius, $\rho$ is constant and we set $T \propto r^{-0.5}$ (the density floor region relaxes to this $T$ profile, so we initialize the models with that profile).  We initialize the specific intensity such that the radiation energy density is $a_rT^4$ (with $T$ the initial gas temperature in the zone) and the angle-integrated radiation flux is $L/(4\pi r^2)$.  For radii less than $R_\star$, $L$ is the luminosity carried by radiation in the \mesa model. Beyond $R_\star$, the gas is optically thin so $L = L_\star$.  The solid lines in panels (a) through (d) of Fig.~\ref{fig:initial_profiles} show, respectively, the initial density, enclosed mass at each radius, (unperturbed) temperature, and luminosity for \ath runs initialized from our $\mzams = 14 \msun$ RSG (\mesa model 14). The grey-shaded region indicates the helium layer in the original \mesa model. 

We drive explosions by initializing an over-pressure region (a ``thermal bomb'') with a total energy of $\edep$ between $\rexp$ and $\rexp + \hexp$. We are interested in explosions that unbind the convective hydrogen envelope of the star, so $\rexp$ is chosen to coincide with the base of the region with $X_{\rm H} > 0.43$ and \texttt{log\_D\_conv} $>$ 15 in the \mesa model. The mass exterior to this radius is $\mconv$.  For each \mesa model chosen for study, we simulate 3 different explosions energies that unbind $\mconv$ with a targeted asymptotic kinetic energy of  $\eej \sim 10^{47}$, $10^{48}$, $10^{49}$, or $10^{50}$ erg. To achieve this, we deposit a total energy of $\edep = \extra + \ebind$ where $\ebind$ is the total binding energy (including thermal) of $\mconv$ and $\extra$ is the targeted asymptotic kinetic energy of the simulation. 

\begin{figure}
\centering
\includegraphics[width=\columnwidth]{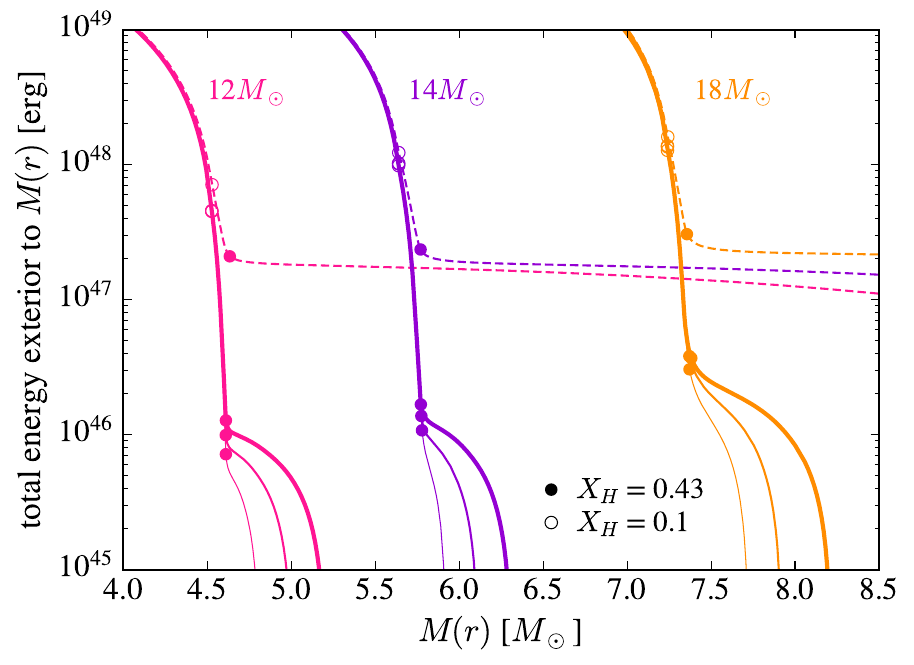}
\caption{Total energy (binding+thermal) of the material exterior to mass coordinate $M(r)$ as a function of mass coordinate for our \mesa progenitors. Dashed curves show the unstripped RSG model for each ZAMS mass (indicated with a label).  For each ZAMS mass, the solid curves show the 3 stripped models in order of increasing envelope mass from left to right. Each filled circle shows the mass coordinate where the hydrogen mass fraction, $X_{\rm H}$ equals 0.43 (the base of the convective H envelope, which is where we place the base of the thermal bomb).  Open circles show where $X_{\rm H} = 0.1$ (near the top of the He core). The difference in energy between these two points is large (a factor of $\sim$several to $>10$) for a very small difference in mass coordinate ($<0.1M_\odot$). Thus, deposited energies can vary widely, yet the explosion will result in essentially fixed $M_{\rm ej}$.}
\label{fig:mesa_energy_vs_mass}
\end{figure}

Our simulations essentially fix $\mej = \mconv$ because $\ebind$ increases steeply inside of $\rexp$. Fig.~\ref{fig:mesa_energy_vs_mass} shows $\ebind$ exterior to $r$ as a function of mass coordinate. The filled circles show $\rexp$ while the open circles mark the outer boundary of the He core.  For the H-poor models (solid lines), there is a factor of $\sim 100$ difference in $\ebind$ between these two radii but with a difference in $M(r)$ of only $\sim0.1\msun$. For the H-rich models (dashed lines) there is a factor of $\sim$few - 10 difference in energy for a difference of $\sim0.1\msun$ in $M(r)$.  Thus, a wide range of energies can be deposited at $\rexp$ in a given \mesa model and result in roughly fixed $\mej$. This is representative of the situation realized in failed SN where accretion of a small amount of mass from the collapsing convective hydrogen envelope drives an explosion of the envelope. The state of the convective material changes with time and radius, so the angular momentum of the accretion flow is uncertain and stochastic. The energy of the explosion powered by this mechanism is, therefore, variable but always larger than the binding energy of the envelope so that $\mej$ is essential fixed for a given stellar model.  Note that this is different from models of supernova where the energy is deposited deeper in and $\mej$ varies with $\edep$. Although $\mej$ is fixed for a given progenitor in our simulations, we probe differences in $\mej$ with similar $\eej$ by instead exploding a suite of \mesa models with varying $\mconv$ from the unstripped models with $\mconv \sim 7-9.2$ $\msun$ to the stripped models with $\mconv\sim0.2-0.6$ $\msun$.

Finally, we note that a complication in mapping the stellar model from \mesa to our \ath setup is that the envelope is convective. The initial luminosity of our \ath model (panel (d) of Fig.\ref{fig:initial_profiles}) is not constant in radius because it only includes the \textit{radiative} luminosity. The convective luminosity in the \mesa model, which is the difference between $L_\star$ and $L_{\rm rad}(r)$ in Fig.~\ref{fig:initial_profiles}, is included in \mesa through an extensive implementation of modified mixing length theory.  We do not include a heating prescription to replace the heat carried by convection.  The luminosity being carried away by radiation from $R_\star$ in our model is, therefore, not being replenished from deeper in. Thus, our model is not in thermal equilibrium. For the weak explosions that we simulate, a thermal wave would propagate inward from $R_\star$ faster than the time it takes for the weak shock to reach the stellar surface. To remedy this, we fix the density and gas pressure to the initial condition in a small region at $\sim$$R_\star$ and set the fluid velocity to zero for $r \gtrsim R_\star$ until the shock arrives.  We have a check on the Mach number and luminosity that detects when the shock approaches the surface. When the shock is detected, we stop rewriting the values. 

\subsection{Grid Resolution}
For most models, our spherical-polar grid extends from an inner radius of 0.1 $\rsun$ to an outer radius of 50,000 $\rsun$. The H-rich, $10^{50}$ erg explosions (12E50, 14E50, 18E50) adopt a larger outer radius of 100,000 $\rsun$ as the ejecta reaches 50,000 $\rsun$ near the end of the plateau. In all models, we adopt a base resolution of 1024 zones, setting \texttt{x1rat} = 1.0107 so that resolution of the base grid is logarithmic in radius.  Additional resolution is needed at larger radii in the stellar envelope and, in particular, in the atmosphere where the density falls of exponentially.  We therefore add 5 levels of nested refinement with the highest refinement region encompassing the outer layers of the star, the exponential atmosphere, and part of the constant density background.  For the H-rich models, the nested refinement region extends from $\sim$ 50 to 20,000$\rsun$ with the highest refinement region between 700 and 3,500 $\rsun$.  The H-poor models are more compact so the nested refinement region is instead placed between 70 and 10,000 $\rsun$ with highest refinement region between 400 and 3,000 $\rsun$. 

In addition, we include 3 levels of refinement on top of the base grid between 1 and 20 $\rsun$ (5 and 40 $\rsun$) for the H-rich (H-poor) models in order to fully resolve the thermal bomb and its conversion from thermal pressure to kinetic energy. To save computational time, this inner region of extra refinement is omitted during later parts of the run when the shock is well outside this region and the gas at those radii comprises the supersonic fallback that we do not care about in the present models. In models with an envelope mass larger than $0.3\msun$, we remap the simulation to a new grid with an inner radius of $1.0\rsun$ after the shock has propagated out to larger radii. This significantly reduces the compute time as the flow through the inner boundary is very supersonic ($\ge$ Mach 10 in the $-r$ direction) and the timestep would be unnecessarily small if we kept the inner boundary at $0.1\rsun$ through the end of the plateau. This remapping takes place well before shock-breakout and, in our tests, has no impact on our results.

\subsection{Summary of Simulations}
\label{secr:parameters}
We simulate a suite of explosions of RSG and YSG stars with $\mzams = 12, 14,$ and $18 \msun$. For each stellar model, we vary $\extra$, which is the energy deposited in the thermal bomb in excess of the binding energy of the envelope. For each unstripped, H-rich star, we simulate $\log(\extra/{\rm erg}) = 48, 49$ and $50$.  For each partially-stripped, H-poor star, we simulate $\log(\extra/{\rm erg}) = 47, 48,$ and $49$.  The suite of \ath simulations is summarized in the first two columns of Table \ref{tab:athmodels}. The \ath model names are as follows.  The number before the E gives the \mesa model from Table \ref{tab:mesamodels} used in the simulation. The first two digits give the ZAMS mass, $\mzams$. For the H-poor models, the decimal number is the mass of the envelope (as defined by \mesans, which is the total mass outside the He core), rounded to one decimal place. The number after the E indicates $\log(\extra/{\rm erg})$.  So 12.3E48 means the \mesa model with $\mzams = 12\msun$ and $\menv \approx 0.3\msun$ was exploded with $\extra = 10^{48}$ erg.  We note that our models focus on explosions that unbind the \textit{inert, convective} hydrogen envelope, $\mconv$.  From Table \ref{tab:mesamodels} we see that \mesa model 12.3 has $\mconv = 0.2\msun$, which is the amount of mass we set up our explosions to unbind. The remaining columns in the table give key results of the simulations, which we discuss in Sec.~\ref{sec:results}.   By simulating explosions of different \mesa models with varying deposited energies, our suite of \ath simulations vary $\mconv$, $\extra$, and $R_\star$ which results in a suite of explosions with varying $\mej$ and $\eej$.

\begin{deluxetable*}{lccccccccccc}
\tabletypesize{\scriptsize}
\tablecaption{Table of \ath simulations.\label{tab:athmodels}}
\tablehead{
Model & $E_{\rm xtra}$ & $M_{\rm ej}$ & $E^{k}_{\rm ej}$ & $v_{\rm ej}$ & $t_{\rm sbo}$ & $L_{\rm peak}$ & $T_{\rm sbo}$ & $\tpl$ & $\lpl$ &$(\lpl
      \tpl)/E_{\rm rec}$\\
	  Name & [$10^{47}$ erg] & $[\msun]$ & [$10^{47}$ erg] & [km/s]  & [hours] & [$10^6 L_\odot$] & [$10^4$K] & [days] & [$10^5 L_\odot$] & \\
}
\startdata 
12E48 &      10 &     6.9 &    14.3 &     342 &      24 &      65 &     5.9 &     300 &      21 &     1.6\\
12E49 &     100 &     6.9 &   172.3 &     983 &      10 &    1095 &     8.3 &     178 &     263 &    12.0\\
12E50 &    1000 &     7.0 &  1379.2 &    3066 &      13 &    4862 &    10.9 &     122 &    1779 &    55.4\\
12.3E47 &       1 &     0.2 &     0.9 &     424 &      30 &      24 &     4.6 &      68 &      11 &     7.7\\
12.3E48 &      10 &     0.2 &    12.4 &    1375 &       8 &     469 &     8.0 &      41 &     130 &    50.2\\
12.3E49 &     100 &     0.2 &   141.1 &    4850 &       7 &    4192 &    15.7 &      26 &    1344 &   308.4\\
12.5E47 &       1 &     0.4 &     0.9 &     320 &      43 &      14 &     4.1 &     104 &      10 &     4.9\\
12.5E48 &      10 &     0.4 &    12.2 &     925 &      12 &     368 &     8.4 &      63 &     112 &    32.0\\
12.5E49 &     100 &     0.4 &   140.3 &    3171 &       7 &    3821 &    12.5 &      39 &    1102 &   192.8\\
12.7E47 &       1 &     0.6 &     0.8 &     269 &      50 &       9 &     4.2 &     134 &       9 &     3.7\\
12.7E48 &      10 &     0.6 &    12.1 &     744 &      17 &     281 &     7.8 &      80 &      99 &    23.7\\
12.7E49 &     100 &     0.6 &   141.2 &    2516 &       9 &    3374 &    12.6 &      50 &     945 &   140.3\\
14E48 &      10 &     7.5 &    16.1 &     345 &      28 &      71 &     5.3 &     315 &      28 &     2.1\\
14E49 &     100 &     7.5 &   187.3 &     951 &      12 &    1243 &     5.2 &     187 &     313 &    14.1\\
14E50 &    1000 &     7.5 &  1433.3 &    2876 &      13 &    5665 &    11.9 &     129 &    1997 &    61.7\\
14.7E47 &       1 &     0.5 &     0.8 &     287 &      58 &      10 &     3.6 &     129 &       9 &     4.1\\
14.7E48 &      10 &     0.6 &    11.0 &     804 &      14 &     309 &     7.0 &      76 &      89 &    23.2\\
14.7E49 &     100 &     0.6 &   137.8 &    2729 &       7 &    3520 &    11.5 &      47 &     881 &   138.5\\
18E48 &      10 &     9.2 &    16.8 &     313 &      35 &      64 &     4.8 &     369 &      33 &     2.5\\
18E49 &     100 &     9.2 &   198.7 &     848 &      14 &    1285 &     8.0 &     214 &     347 &    15.1\\
18E50 &    1000 &     9.3 &  1518.8 &    2507 &      14 &    6449 &    11.9 &     148 &    2160 &    64.3\\
\enddata
\tablecomments{\\Model name -- The base model name indicates the \mesa progenitor from Table 1. For the \ath runs, we append E with the explosion energy. E.g. E49 means $E_{\rm xtra} = 10^{49}$ erg is deposited in addition to $|\ebind|$ from Table 1. \\
	$M_{\rm ej}$ -- Ejecta mass. \\ 
        $E_{\rm ej}$ -- Asymptotic kinetic energy of the ejecta. \\
        $v_{\rm ej}$ -- Asymptotic velocity of the ejecta. \\
        $t_{\rm sbo}$ -- FWHM of the shock cooling peak.\\
	$L_{\rm peak}$ -- Maximum luminosity of the lightcurve. \\
        $T_{\rm sbo}$ -- Photosphere temperature at peak. \\
	$\tpl$ -- Total duration of the event from shock breakout and until the end of the plateau. \\ 
        $\lpl$ -- Luminosity at the midpoint of the plateau.\\
        $(\lpl\tpl)/E_{\rm rec}$ -- Roughly the ratio of radiated energy on the plateau to the energy available from hydrogen and helium recombination of the ejecta. }
\end{deluxetable*}

\section{Results and Analysis}
\label{sec:results}
In this section, we discuss the results of our \ath simulations summarized in Table \ref{tab:athmodels}.  We first discuss the dynamical properties of the explosions, then we discuss the resultant light curves.

\subsection{Hydrodynamical evolution of the shock and ejecta}
\begin{figure*}
\centering
\includegraphics[width=1.0\textwidth]{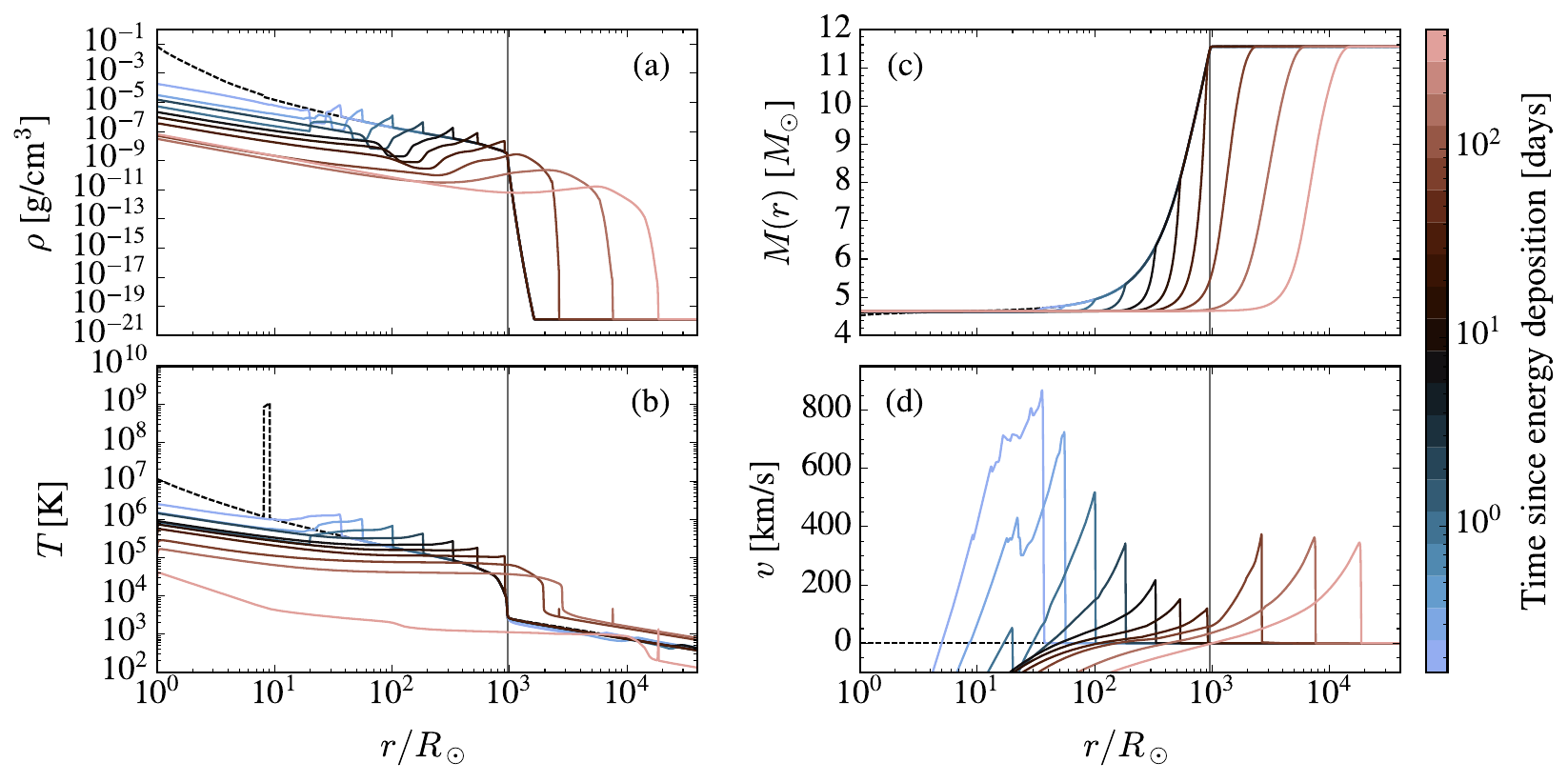}
\caption{Evolution of the shock and ejecta after deposition of $E_{\rm xtra} = 10^{48}$ erg into the envelope of the H-rich model with $\mzams = 12\msun$ and $\mconv = 6.9 \msun$ (model 12E48). The thermal pressure bomb (dashed peak in the temperature panel) launches a pressure wave that steepens into a shock and lifts material beyond the stellar surface (vertical solid line in all panels).}
\label{fig:hrich_times_series}
\end{figure*}

Figure \ref{fig:hrich_times_series} presents radial profiles of the hydrodynamical properties of model 12E48 following deposition of the thermal bomb.  The dashed, black curve shows the initial condition while the remaining curves are approximately logarithmically spaced in time.  Panel (a) is gas density, panel (b) is the gas temperature, panel (c) is the mass enclosed at each radius, $r$, and panel (d) is the velocity profile of the gas. The vertical line is the stellar surface.  The density and temperature panels show the evolution of the over-pressure region into an outgoing shock wave that propagates through the envelope.  The temperature behind the shock is roughly constant as it approaches the surface. After shock breakout, material is lifted beyond the surface of the star.  At late times, the photosphere (mediated by a recombination wave) has propagated inward in mass coordinates to the inner boundary as evidenced by the final temperature profile showing that the material on the domain has cooled.  The velocity panel shows that the shock starts out with a large velocity that declines as the shock sweeps up envelope mass until it reaches the surface. Shock breakout accelerates the shock to high velocities and by the last two snapshots the shock velocity (the maximum of $v(r)$ which is the same as the velocity of the shock front in the lab frame) is constant at $\approx$340 km s$^{-1}$.  

\begin{figure*}
\centering
\includegraphics[width=1.0\textwidth]{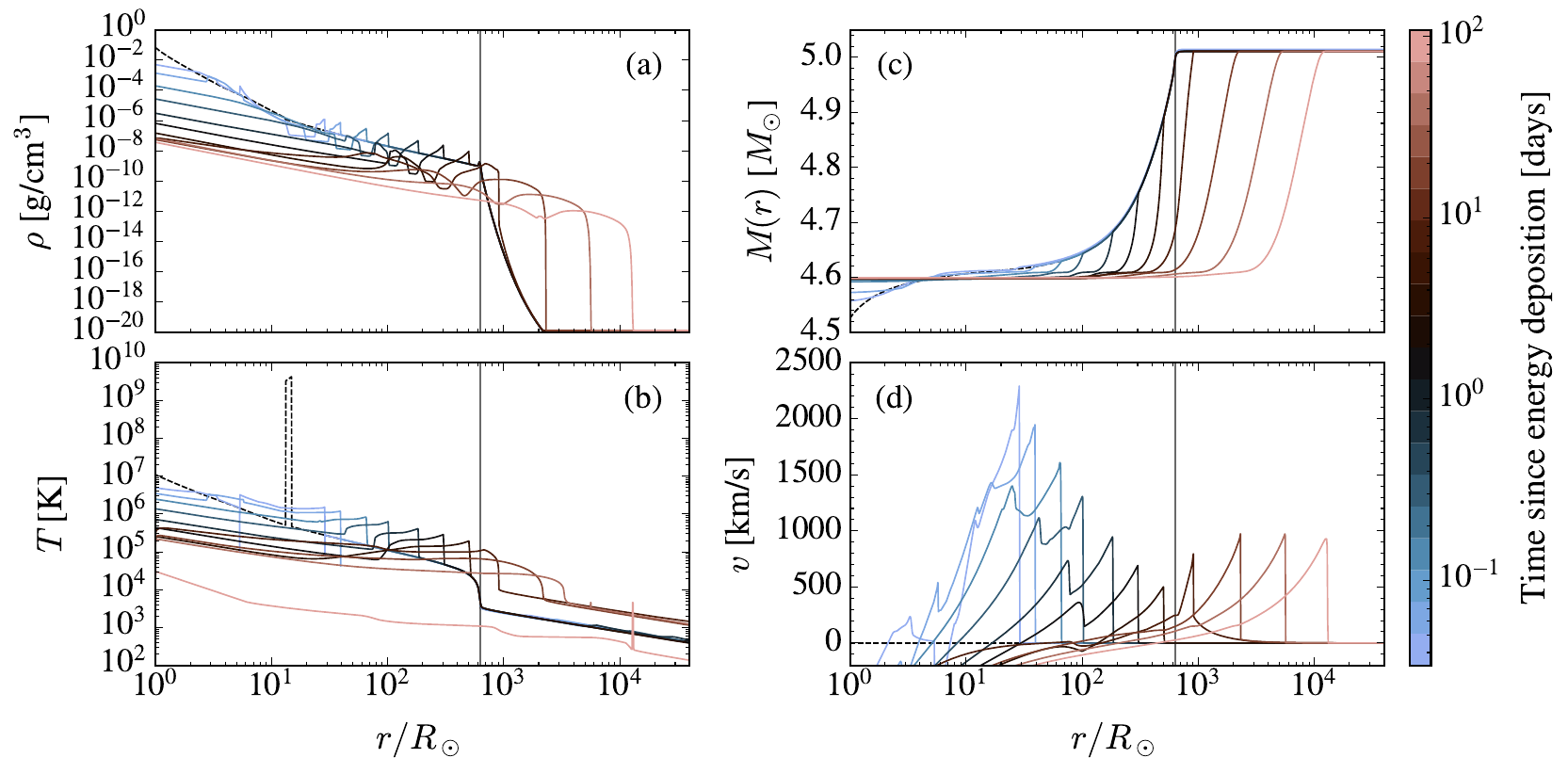}
\caption{Evolution of the shock and ejecta after deposition of $E_{\rm xtra} = 10^{48}$ erg into the envelope of the H-poor model with $\mzams = 12\msun$ and $\mconv = 0.40\msun$ (model 12.5E48). The evolution is similar to that of Fig.~\ref{fig:hrich_times_series} but the velocities are several times larger owing to the lower envelope mass of the star.}
\label{fig:hpoor_times_series}
\end{figure*}

Figure \ref{fig:hpoor_times_series} shows a similar figure but for model 12.5e48, which is analogous to model 12E48 ($\mzams = 12\msun$ and $\extra = 10^{48}$) except that the model has an envelope mass of $\mconv = 0.4\msun$ (rather than $6.9\msun$ as in model 12E48). The hydrodynamical evolution of the shock is very similar, except that everything occurs faster owing to the lower envelope mass (note the difference in time range on the color bar axis). In addition, because $\extra$ is the same for lower $\mconv$, the shock velocities are always faster.  The leading edge of the shock asymptotes to $v \approx 925$ km s$^{-1}$, roughly a factor of 3 larger than in the 12E48 model, consistent with $\vej\propto \sqrt{\extra/\mconv}$. 

Finally, the velocity profiles in panel (d) of Figs. \ref{fig:hrich_times_series} and \ref{fig:hpoor_times_series} show two peaks at early times due to the spreading of the thermal bomb in radius (and thus spanning a range of binding energies). This feature is more pronounced and persists through shock breakout in the H-poor model due to the lower mass of the envelope that the shock must sweep through.  Notice, however, that following shock breakout the leading peak accelerates and the trailing peak falls back so that the leading feature of the shock is the feature that gives rise to the ejecta and the asymptotic behavior of the energetics.


\begin{figure}
\centering
\includegraphics[width=1.0\columnwidth]{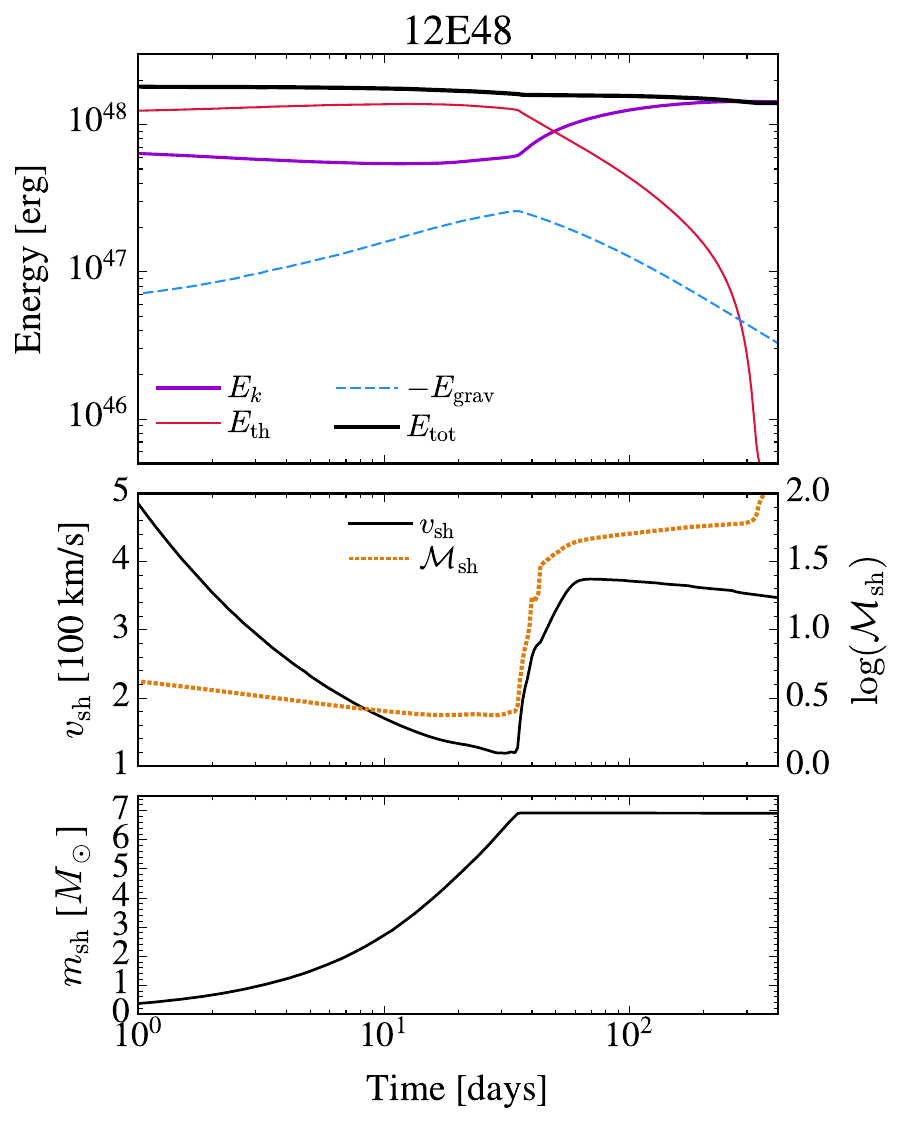}
\caption{Shock and ejecta properties as a function of time for model 12E48.}
\label{fig:shock_12E48}
\end{figure}

To compare ejecta properties across \ath models, we compute asymptotic properties of the ejecta by following the mass and energies entrained in the shock over time.  The leading edge of the shock is the outermost local maximum in the velocity profile.  We define the trailing edge as the point where the velocity turns negative.  Within this region, we measure the total mass, kinetic energy, thermal energies, and gravitational binding energy. The upper panel of Fig.~\ref{fig:shock_12E48} shows the energies as a function of time for the representative case, model 12E48 (the same model shown in Fig.~\ref{fig:hrich_times_series}). The middle panel shows the shock velocity (solid, black line; left $y$-axis) and the shock Mach number (dotted, gold line; right $y$-axis).  The bottom panel shows the mass entrained in the shock (and outflow).  As we saw in Fig.~\ref{fig:hrich_times_series}, the shock velocity declines up to day $\sim$30 as the shock gains mass while sweeping through the stellar envelope. While in the envelope, the shock Mach number is of order a few.  After day 30, the shocked ejecta has exited the star and the mass entrained in the shock has become constant.  After an initial acceleration associated with shock breakout, the shock velocity becomes roughly constant at $\approx 340$ km s$^{-1}$ (though slowly declining). The top panel shows that after day $\sim 200$ the kinetic energy of the ejecta is roughly equal to the total energy.  The thermal energy in the ejecta steadily declines after day 30 as the photosphere sweeps inwards in mass coordinates and material outside the photosphere cools.  The total energy in the shock and ejecta is always positive because we have deposited $\extra =10^{48}$ erg in addition to the thermal and gravitational binding energy of the convective envelope.  Notice that the kinetic energy asymptotes to $\eej\sim \extra$, by construction.  

For each \ath model, we take the mass, kinetic energy, and velocity of the ejecta ($\mej$, $\eej$, and $\vej$, respectively) to be the asymptotic values of those quantities in calculations like those in Fig.~\ref{fig:shock_12E48}. The results for our suite of simulations are listed in Table~\ref{tab:athmodels}.  Referring to the fifth column of the table, for the more massive, RSG envelopes, the lower energy explosions ($10^{48}-10^{49}$ erg) result in ejecta velocities of $\sim 300 - 1000$ km s$^{-1}$. A $10^{50}$ erg explosion is needed to push $\vej$ above 1000 km s$^{-1}$. Due to their lower envelope masses, explosions of the YSGs at the same energies necessarily gives rise to higher ejecta velocities. For example, RSG model 12E49 has $\vej$ of 983 km s$^{-1}$ but the YSG model 12.5E49 achieves $\vej \approx 3200$ km s$^{-1}$.

\subsection{Light Curves}
\begin{figure*}
\centering
\includegraphics[width=0.9\textwidth]{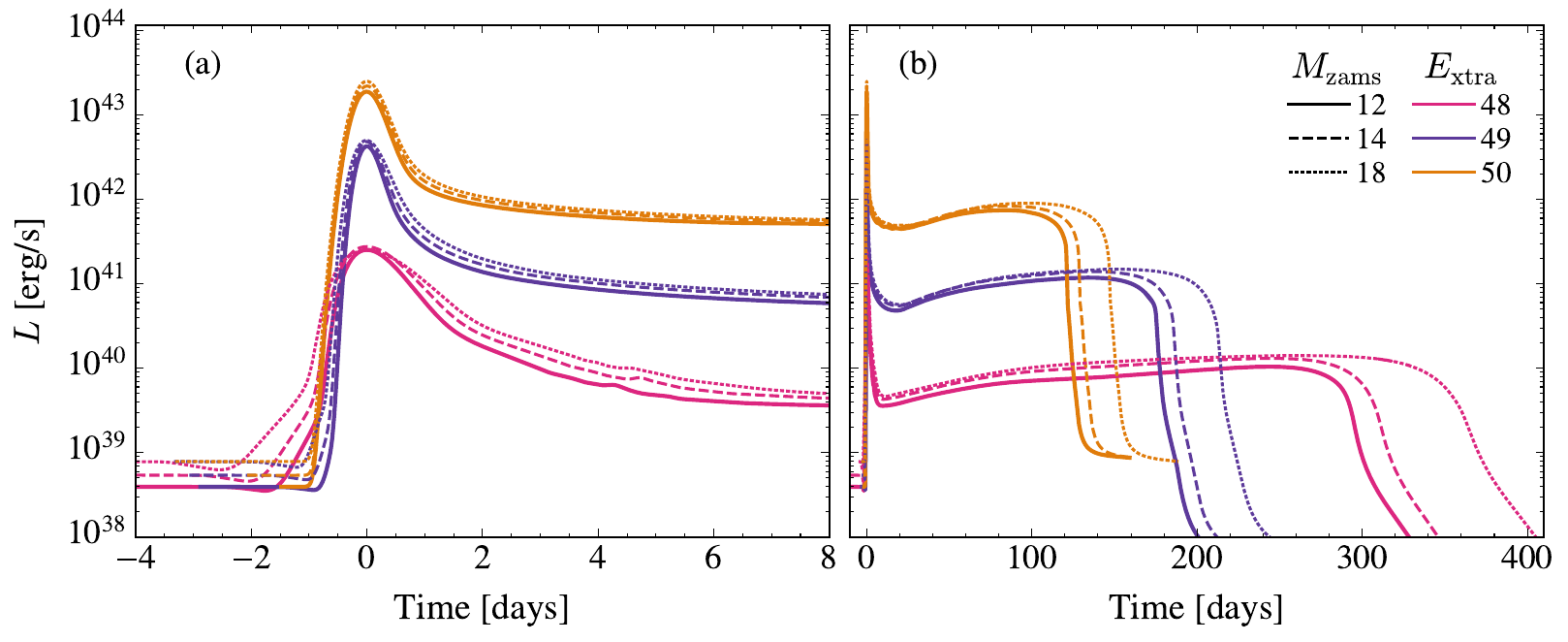}
\caption{Lightcurves of explosion models of the H-rich, RSG progenitors as a function of time. The lightcurves have been shifted to the same phase such that peak brightness is at time = 0.  Data is the same in both panels; the left panel zooms-in on the shock breakout peak. Line style indicates the ZAMS mass of the star. Color indicates energy deposited in addition to the binding energy of the convective envelope, which is roughly the asymptotic kinetic energy of the ejecta.}
\label{fig:lightcurves_hrich}
\end{figure*}

We now present the observable consequences of the explosions we have simulated.  Fig.~\ref{fig:lightcurves_hrich} shows the bolometric luminosity versus time for all of the H-rich explosions. The left panel is restricted in time to highlight shock breakout while the right panel shows the full light curves. The lower limit of the y-axis is just below the luminosity of the progenitors ($\approx 4-9 \times 10^{38}$erg s$^{-1}$). The qualitative features of the lightcurves are what we would expect from ejection of hydrogen-rich material from the stellar surface. There is an initial peak due to breakout of the shock from the surface of the star and then rapid cooling of material near the photosphere.  The shock breakout / cooling (SBO) peak transitions to a long plateau in the lightcurve as a hydrogen recombination front (and thus the photosphere) recedes in mass coordinates through the expanding ejecta.  

These features are reminiscent of SNe IIp lightcurves though with important differences.  First, the lightcurves are scaled down in luminosity because significantly less thermal energy was deposited in the envelope by the shock versus the $\sim 10^{51}$ erg explosion energy of a IIp.  Second, the durations of the events are longer than the typical $\sim$100 day plateau of SN IIp lightcurves because the ejecta velocities or $\sim 100 - 1000$ km s$^{-1}$ instead of 4,000 km s$^{-1}$, as in IIps.   Another critical difference is that there is no explosive nucleosynthesis in low-energy ejections of a mostly hydrogen layer so our lightcurves do not transition to an exponential tail from radioactive decay at the end of the plateau. Instead, the luminosity drops sharply at the end of the plateau in our models.

\begin{figure*}
\centering
\includegraphics[width=0.9\textwidth]{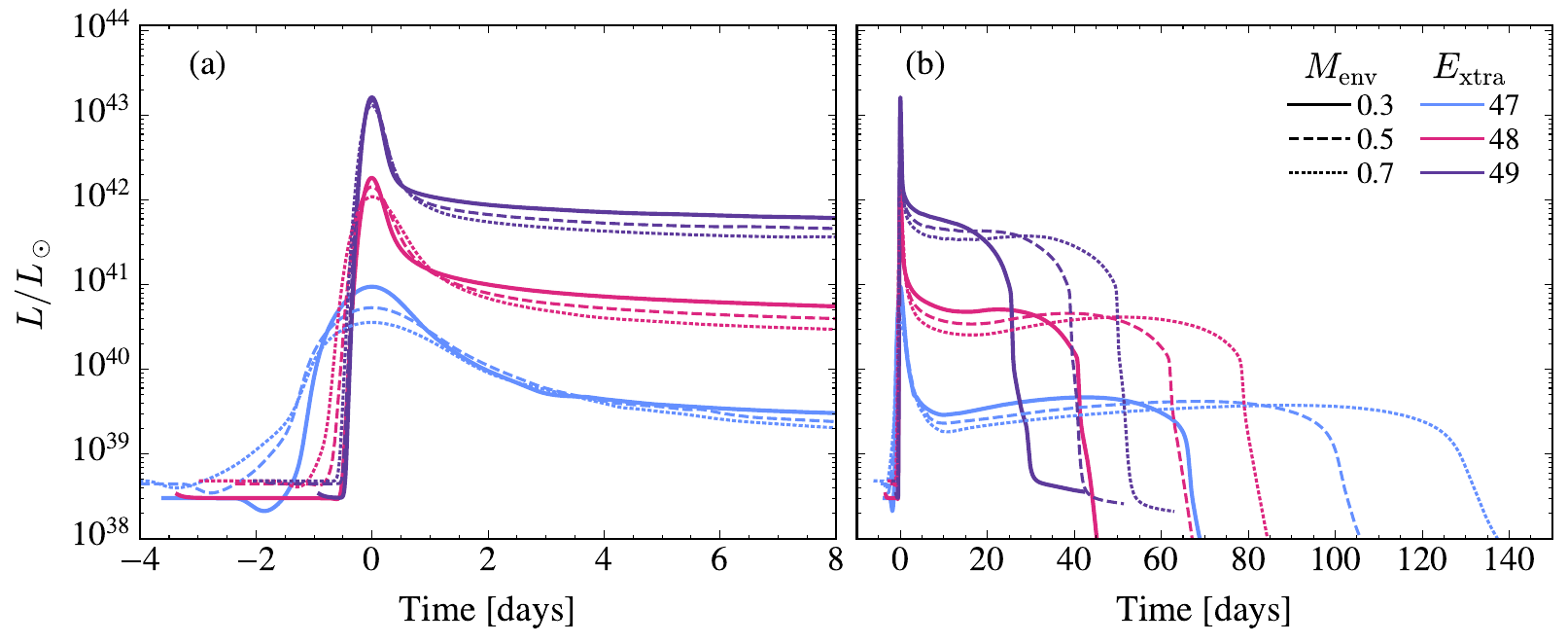}
\caption{Same as in Fig.~\ref{fig:lightcurves_hrich} but showing explosions of the H-poor, YSGs with $\mzams = 12\msun$. Here line style refers to the envelope mass as given in the \mesa model name.}
\label{fig:lightcurves_hpoor}
\end{figure*}

Fig.~\ref{fig:lightcurves_hpoor} is analogous to Fig.~\ref{fig:lightcurves_hrich} but instead shows the lightcurves for our explosions of H-poor stars. Qualitatively, the lightcurves show the same features as the H-rich explosions except have overall shorter durations and higher luminosities owing to the lighter envelope masses for similar explosion energies.

\begin{figure*}
\centering
\includegraphics[width=1.0\textwidth]{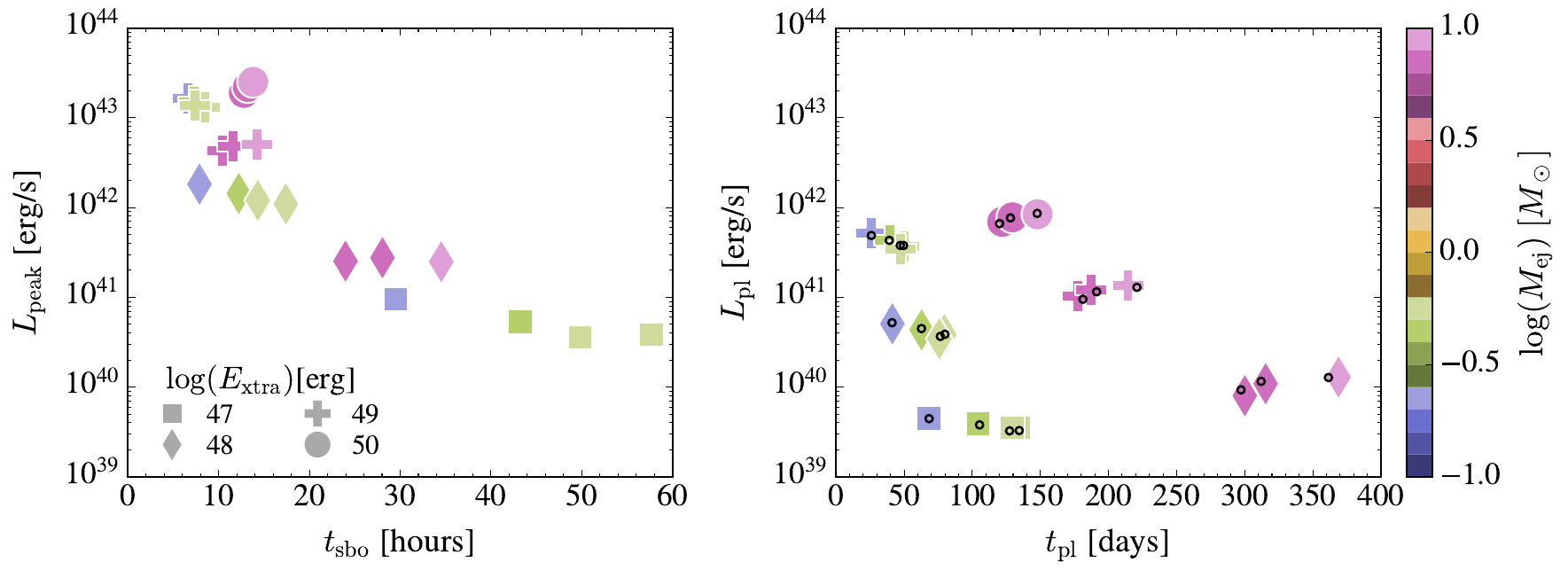}
\caption{Characteristic luminosities and timescales of the lightcurves for all models. \textit{Left panel}:  Peak brightness versus duration of the shock breakout / shock cooling (collectively, SBO) bump where duration is measured as the FWHM of the SBO feature.  \textit{Right panel}: Luminosity on the plateau and the total duration of the event from rise above the stellar luminosity to end of the plateau. Symbol shape indicates the deposited extra energy while symbol color indicates the ejecta mass. The small circles show our fits to $\lpl$ and $\tpl$ (eqs.~\eqref{eq:lpl_fit} and \eqref{eq:tpl_fit}).}
\label{fig:L_t_peaks}
\end{figure*}

We quantify the two characteristic features of the lightcurves, the SBO peak and the plateau, by measuring their characteristic luminosities and durations. 
For the SBO peak, we measure the maximum bolometric luminosity, $L_{\rm peak}$, and take the duration, $t_{\rm sbo}$, to be the full-width-half-maximum of the SBO feature. For the plateau, we take the duration, $\tpl$, to be the full duration of the lightcurve from the rise to maximum through the end of the plateau (when the luminosity has fallen to $\approx 10\%$ of the peak luminosity. The luminosity on the plateau, $\lpl$, is measured at $t = \tpl/2$. In practice this ends up being the same as the time-weighted average luminosity over the duration of the light curve.  The characteristic luminosities and durations are listed in Table~\ref{tab:athmodels} and are plotted in Fig.~\ref{fig:L_t_peaks}.  
We now discuss the SBO and the plateau phases in turn.

\subsection{Shock breakout / shock cooling}
\label{sec:sbo}
Characteristic SBO quantities are shown in the left panel of Fig.~\ref{fig:L_t_peaks} which plots $L_{\rm peak}$ versus $t_{\rm sbo}$ (in hours) as a function of ejecta mass (given in the color bar). The H-poor YSGs are the blue and green symbols while the H-rich RSGs are in shades of pink.  Symbol shape indicates $\extra$ for each model. SBO durations range from $\sim 5-10$ hours for the largest energy explosions while lower energy explosions have shock breakout signatures lasting 10s of hours.  SBO is sensitive to stellar radius; breakout of $10^{49}$ erg explosions of our YSG models is $\approx 6-8$ hours while a breakout from our RSGs with similar explosion energies lasts $\approx$10-14 hours.  

The SBO durations quoted here are surely sensitive to the density profile at the surface of the star.  The density at the surface of our \mesa models is sensitive to the choices that we made in our stellar modeling, namely in our choice of handling super-adiabatic convection. The prescription for super-adiabatic convection effects both the shape of the density inversion at the surface (which is itself an unphysical artifact of solving the stellar structure equations in 1D) as well as the slope of the density profile interior to the density inversion. When mapping the models to \athns, we paste on an exponential, isothermal atmosphere.  A more shallow density profile would result in longer breakout signatures \citep[e.g.,][]{Lovegrove2017}. Additionally, RSGs and YSGs also have transonic surface convection that results in density variations of a factor of $\sim$10 \citep{2022ApJ...929..156G}. As a result, the optical depth at which breakout occurs ($\tau \sim c/v_{\rm sh}$) is located at a different radius at each location in the turbulent structure.  For IIp explosions, \citet{2022ApJ...933..164G} found that the convective structure smeared out the 1.5 hour SBO signature of their fiducial $8 \times 10^{50}$ erg explosion to 6 hours.  Following \citealt{2022ApJ...933..164G} we can estimate whether this effect is important at the energies we simulate as follows.  The 3D structure is important for setting the SBO duration if the time it takes the shock to cross the extended, clumpy layer introduced by the 3D convection is longer than the diffusion time through the $\tau < c/v_{\rm sh}$ layer. The former is $t_{\rm cross} = \Delta R / v_{\rm sh}$ with $\Delta R$ the thickness of the clumpy transonic convective layer ($\Delta R \sim 80 - 100$ $\rsun$ in the 3D envelopes of \citealt{2022ApJ...929..156G}). The latter is $t_{\rm diff} \sim c / (\kappa \rho v_{\rm sh}^2)$ with $\kappa$ and $\rho$ measured at $\tau \sim c/v_{\rm sh}$ near breakout.  Thus, $t_{\rm cross} > t_{\rm diff}$ when $v_{\rm sh} > c / (\kappa \rho \Delta R)$. In our $10^{48}$ erg RSG explosions, $\kappa \approx 1$ cm$^2$/g  and $\rho \approx 10^{-8.5} - 10^{-8}$ g/cm$^3$ at $\tau \sim c/v_{\rm sh}$. For those values, $t_{\rm cross} > t_{\rm diff}$ when $v_{\rm sh} \gtrsim 5$ km s$^{-1}$. In our models, $v_{\rm sh} \gtrsim 100$ km s$^{-1}$ just before shock breakout, so the duration of the SBO signature would, indeed, be set by $t_{\rm cross}$. That is, the 3D convective structure sets $t_{\rm sbo}$, which is not captured in our 1D models.  We also do not correct $t_{\rm sbo}$ for the finite light travel time across the stellar disc, $R_\star / c$.  For a star with $R_\star = 1000 \rsun$, $R_\star / c \approx 0.6$ hours.  This is smaller than the other uncertainties in our models, so we do not correct for that effect (incidentally, the light travel time is much shorter than $t_{\rm cross}$; the 3D structure sets the SBO duration in RSGs and, likely, YSGs).  Our primary interest in this paper is in the long term evolution of the lightcurve (i.e. the plateau); we have reported the shock breakout signatures here for completeness.

\subsection{Long term evolution (the plateau)}
The right panel of Fig.~\ref{fig:L_t_peaks} shows the luminosity on the plateau versus the lightcurve duration as a function of ejecta mass for our suite of explosion simulations. Comparing to the left panel, the luminosity on the plateau is roughly a factor of 10 lower than the SBO peak (though the difference is larger at larger energies and lower ejecta masses).   Our H-poor models (blue and green symbols) have durations of $\sim$25 to 140 days, while the H-rich explosions (pink symbols) have durations of $\sim$125 to 370 days. The H-poor explosions are shorter and brighter than H-rich explosions of similar energies because lower $\mej$ at fixed $\eej$ implies larger $\vej$ ($\eej \sim \frac{1}{2}\mej\vej^2$).  Larger $\vej$ means faster evolution (i.e. shorter duration) as well as higher luminosity because the thermal energy deposited by the shock must be radiated over a shorter time (i.e. $\lpl\propto\eej/\tpl$). 

As in SNe IIP, the lightcurve plateau in our models arises due to hydrogen recombination in the expanding ejecta. As the ejecta expands and cools, the hydrogen recombines from the outside in, resulting in a dramatic drop in the opacity due to the loss of free electrons in the now neutral outer layers. In SN IIp ejecta, the gas is radiation-pressure dominated and the dominant opacity source behind the recombination front is from electron scattering. Adopting those assumptions and assuming the thermal energy deposited by the shock is the dominant energy source (i.e. neglecting gravity, recombination energy, and the thermal energy of the unperturbed envelope), \citet{Popov1993} used a one-zone model to find scalings for $\lpl$ and $\tpl$. The ejecta is considered uniform except that there is an opacity jump at the recombination front.  Inside the recombination front, the opacity is assumed constant (equal to the Thompson opacity). Exterior to the recombination front, the material is assumed to have zero opacity. The luminosity on the plateau is then estimated as $\lpl = 4\pi\sigma R_{\rm rec}^2 T_I^4$ where $R_{\rm rec}$ is the radius of the recombination front and $T_I$ is the ionization temperature. The duration of the event, $\tpl$, can be estimated by computing the time that it takes for $R_{\rm rec}$ to sweep inward through the hydrogen-rich ejecta.  \citet{Popov1993} found 
\begin{align}
    \lpl &= 1.64 \times 10^{42} \frac{R_{500}^{2/3} E_{51}^{5/6} T_{5054}^{4/3}}{M_{10}^{1/2}\kappa^{1/3}_{0.34}} {\rm erg/s},
    \label{eq:lpl_popov}
\end{align}
and
\begin{equation}
    \tpl = 99\frac{\kappa_{0.34}^{1/6}M_{10}^{1/2}{R_{500}^{1/6}}}{E_{51}^{1/6}T_{5054}^{2/3}} {\rm days}
    \label{eq:tpl_popov}
\end{equation}
where $M_{10} =\mej/(10\msun)$, $E_{51}=\eej/(10^{51} {\rm erg})$, $R_{500}=R_\star/(500\rsun)$, $\kappa_{0.34} = \kappa/(0.34 {\rm cm}^2/{\rm g})$, and $T_{5054} = T_I/(5054 {\rm K})$ \citep[see also][]{1980ApJ...237..541A}.   
Similar scalings for SNe IIp have been obtained numerically by \citet{Kasen_2009} and \citet{Goldberg2019}.  Our $10^{50}$ erg RSG explosions are consistent with the \citealt{Kasen_2009} and \citealt{Goldberg2019} scaling but their numerical fits do not do as good a job of modeling our lower energy or lower envelope mass results.  Instead, we find the original \citealt{Popov1993} scaling are more accurate than the \citealt{Kasen_2009} and \citealt{Goldberg2019} fits for our lower energy and lower ejecta mass models. 

The \citealt{Popov1993} scalings with $T_I = 5054$ K and $\kappa =$ 0.3 or 0.33 (the scattering opacity with $X_{\rm H} =$ 0.5 or 0.65, respectively, as adopted in our simulations) do a good job of describing our H-poor simulations as well as our H-rich models with $\extra \ge 10^{49}$ erg. For the H-rich, $\extra = 10^{48}$ models (12E48, 14E48, 18E48), $\lpl$ is predicted well but $\tpl$ is predicted to be $\sim 30-50$ days shorter than our simulated explosions. This can be corrected by (arbitrarily) setting $\kappa=0.6$ (this is still lower than the opacity inside the emitting region in our simulations; see below). We caution against applying the \citealt{Popov1993} scalings to RSG explosions below $\eej \sim 10^{48}$ erg and to YSG explosions below $\eej \sim 10^{47}$ erg without correcting for differences in $\kappa$ (and, potentially, $T_I$). The emitting region on the plateau transitions to becoming gas-pressure dominated at those energies and, among other things, the opacity is enhanced above pure electron scattering (see below). Indeed, the \citealt{Popov1993} scalings assume radiation-pressure dominated gas so will necessarily not apply below the energy cutoffs we have given. 

For completeness, and because one may not know the opacity or ionization temperature to assume, we provide the following numerical fits to our models without $\kappa$ or $T_I$ dependence, thus absorbing those dependencies into the scaling with $\mej$, $\eej$, and $R_\star$. Using the data from the right panel of Fig.~\ref{fig:L_t_peaks}, we find
\begin{align}
    \lpl =  2.46 \times 10^{38}\bigg(\frac{\mej}{10\msun}\bigg)^{-0.80}\bigg(\frac{\eej}{10^{47} \rm erg}\bigg)^{0.94} \nonumber\\
    \times \bigg(\frac{R_\star}{500\rsun}\bigg)^{1.3} {\rm erg/s}, 
    \label{eq:lpl_fit}
\end{align}
and
\begin{align}
    \tpl = 521 \bigg(\frac{\mej}{10\msun}\bigg)^{0.53}\bigg(\frac{\eej}{10^{47} \rm erg}\bigg)^{-0.20} \nonumber \\
    \times\bigg(\frac{R_\star}{500\rsun}\bigg)^{0.25} {\rm days}.
    \label{eq:tpl_fit}
\end{align}
These fits are plotted with small black circles in the right panel of Fig.~\ref{fig:L_t_peaks}.  The sign of the powers in our fits are the same as in eqs.~\eqref{eq:lpl_popov} and \eqref{eq:tpl_popov} but $\lpl$ has a stronger dependence on $\mej$ and $R_\star$ and $\tpl$ a stronger dependence on the $R_\star$. 

\begin{figure*}
\centering
\includegraphics[width=1\textwidth]{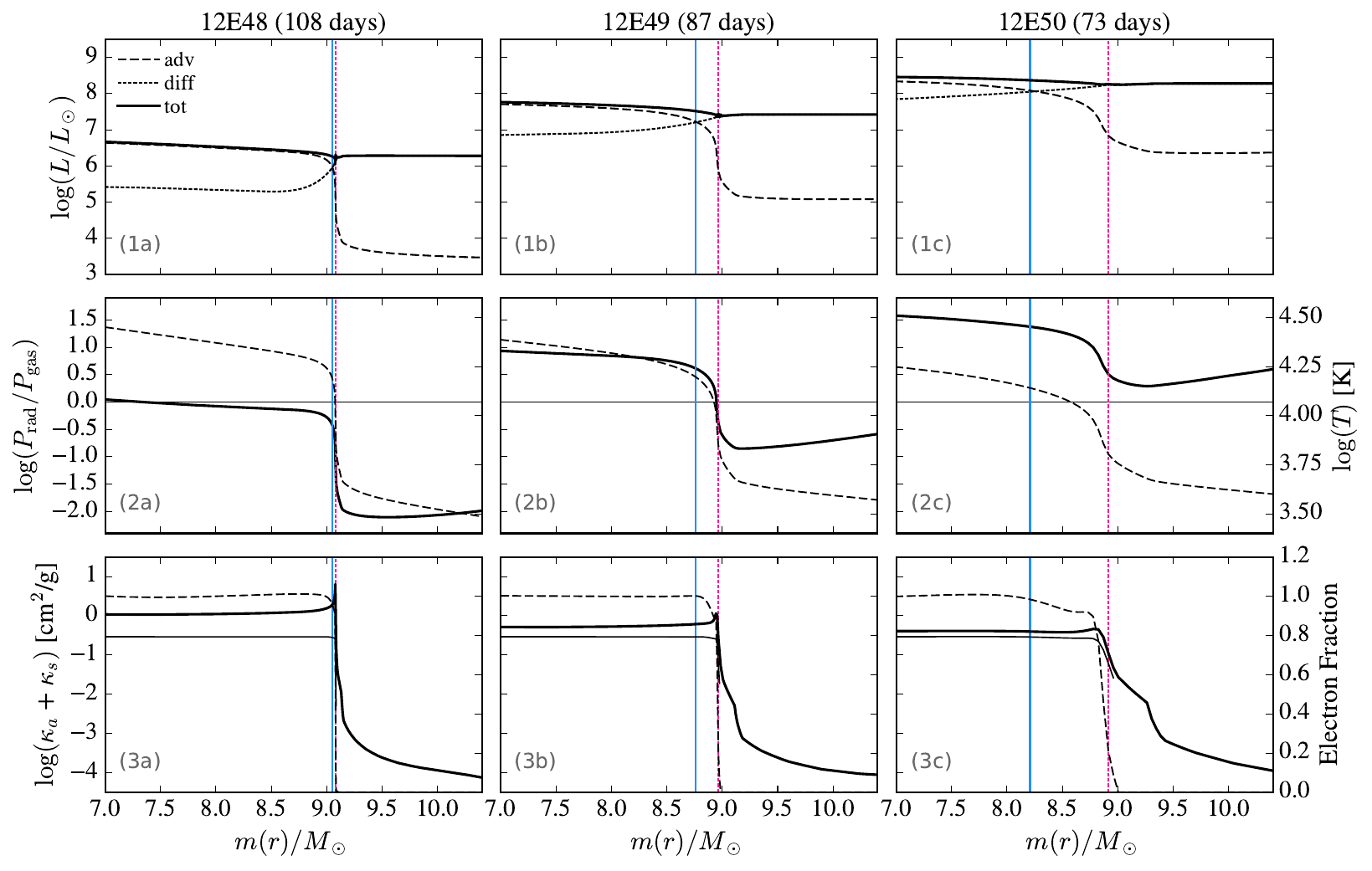}
\caption{Profiles versus enclosed mass highlighting properties of the emitting region on the plateau for models with $\mzams = 12\msun$. The three columns show, from left to right, a snapshot from the model with $\log(\extra/{\rm erg}) = 48, 49,$ and $50$, respectively. The title of each column gives the time in days since the peak of the lightcurve (as in Fig.~\ref{fig:lightcurves_hrich}). \textit{Top row:} Total luminosity (solid line), diffusive luminosity (dotted line), and advective luminosity (dashed line). \textit{Second row}: Ratio of radiation pressure to gas pressure (solid line, left $y$-axis) and gas temperature (dashed line, right $y$-axis). \textit{Third row}: Total Rossland mean opacity (including absorption and scattering; thick solid line, left $y$-axis), scattering opacity (thin solid line, left $y$-axis), and electron fraction (dashed line, right $y$-axis). In each panel, the vertical blue line marks the mass coordinates where $L_{\rm diff} = L_{\rm adv}$ and the vertical pink, dashed line is where $L_{\rm diff}\approx L_{\rm tot}$ (the last scattering surface).}
\label{fig:emit_layer_12}
\end{figure*}

We illustrate in Fig.~\ref{fig:emit_layer_12} how the conditions in the emitting region on the plateau differ with explosion energy in the H-rich explosions. Each column of the figure corresponds to a snapshot in time (during the plateau) for models with fixed progenitor (the RSG model with $\mzams = 12\msun$ and $\mconv=6.9\msun$) but different $\extra$.  We choose snapshots such that the recombination front has reached a similar mass coordinate in all three simulations.  Each column is labeled with the time of the snapshot relative to the lightcurve maximum (same as the $x-$axis in Fig.\ref{fig:lightcurves_hrich}).  
The first row shows the different contributions to the luminosity as a function of mass. The solid line is the total luminosity, the dashed line is just the advective component, and the dotted line is the diffusive component.  
In the second row, the solid line shows the log of the ratio of gas to radiation pressure (the horizontal line at 0 is where gas and radiation pressure are equal).  The dashed line is the log of the gas temperature.  The final row shows the total absorption and scatting opacity (thick solid line) and the opacity from scattering alone (thin solid line).  The dashed line is the electron fraction (average number of free electrons per atom).  In each panel of the figure, the pink, dotted vertical line indicates the mass coordinate of the photosphere where the optical depth is $\tau_{\rm ph} \approx 1$. The blue, solid line is the trapping radius, where $\tau = \tau_{\rm crit} \equiv P_{\rm rad} (c/ v_{\rm sh})/(P_{\rm rad} + P_{\rm gas})$\footnote{In practice, we compute these two locations as the locations where $L_{\rm diff} \approx L_{\rm tot}$ and where $L_{\rm diff} = L_{\rm adv}$, respectively. If we integrate $\tau$ from the outer radius of our domain, we would integrate over a region where some molecular opacity increases the value of the integral artificially (and shifts, e.g., the mass coordinate of $\tau=1$, even though this region does not impact the radiated luminosity from the ejecta).}.  

The second row of Fig.~\ref{fig:emit_layer_12} shows that gas pressure becomes increasing important at lower explosion energies.  
While the emitting region of the $10^{50}$ erg explosion is very radiation-pressure dominated (panel 2c), gas pressure is slightly dominant at $\tau_{\rm crit}$ in the $10^{48}$ erg explosion and $P_{\rm gas} \approx P_{\rm rad}$ deeper in (panel 2a).  In addition, the solid lines in the final row of the figure show that, as the energy decreases, the total opacity (thick solid line) behind the recombination front (the step in the dashed line) increases above the pure scattering opacity (thin horizontal line) as absorption opacities (e.g. free-free / inverse Bremsstrahlung) become more important. 

\begin{figure*}
\centering
\includegraphics[width=1\textwidth]{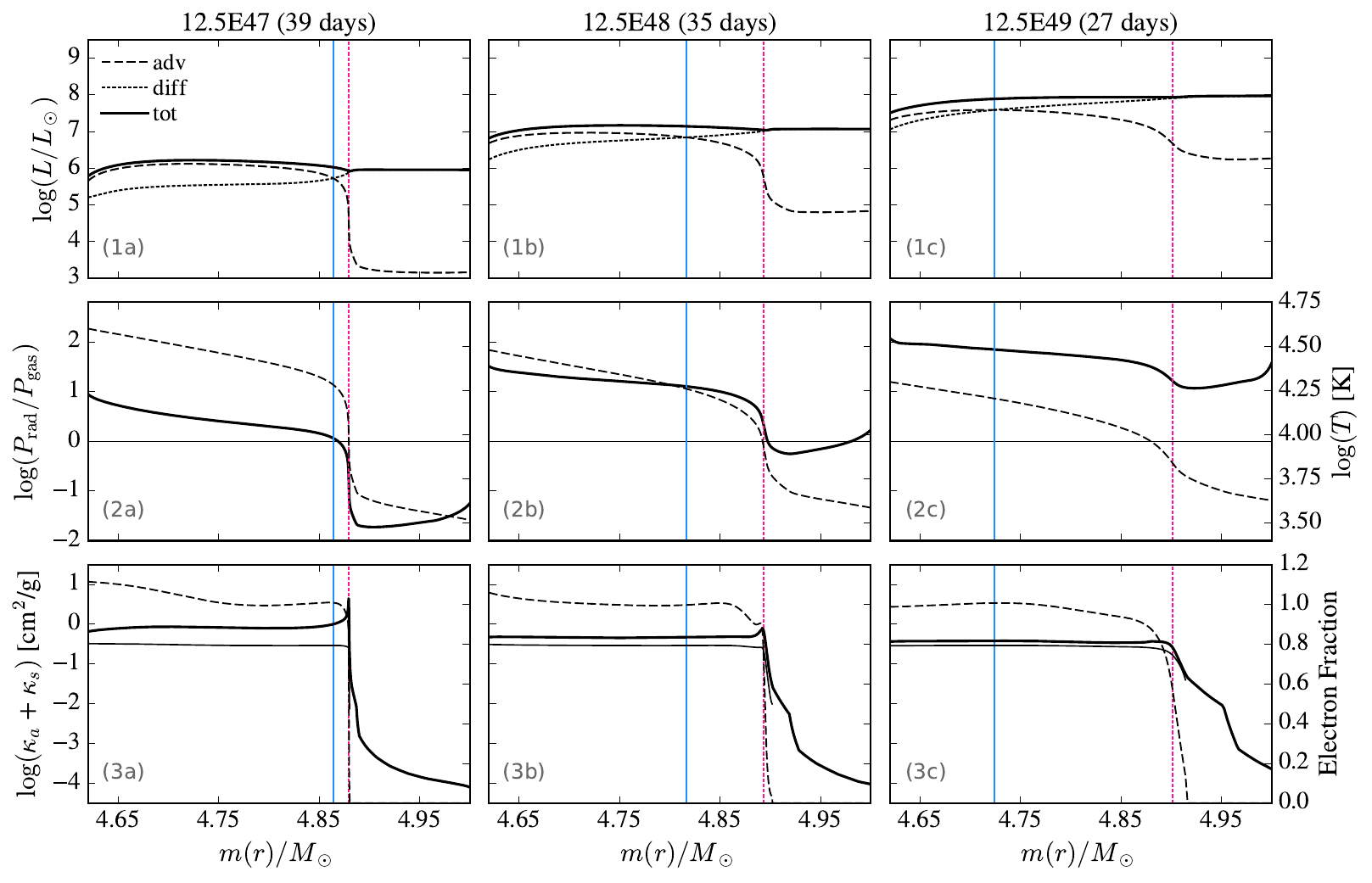}
\caption{Same as Fig.~\ref{fig:emit_layer_12} but for the hydrogen poor models with $\mzams = 12\msun$ and $\mconv = 0.4\msun$. The three columns show, from left to right, a snapshot from the model with $\log(\extra /{\rm erg}) = 47, 48,$ and $49$, respectively.}
\label{fig:emit_layer_12.5}
\end{figure*}

Fig.~\ref{fig:emit_layer_12.5} is analogous to Fig.~\ref{fig:emit_layer_12} but instead shows explosions of the YSG with $\mzams = 12\msun$ and $\mconv=0.4\msun$. For these lower-$\mej$ models, the transition to the gas-pressure-dominated regime happens near $\eej \sim 10^{47}$ erg instead of $\sim10^{48}$ erg.  

\section{Discussion}
\label{sec:discussion}

\subsection{Comparison to previous studies}
Three previous studies have computed lightcurves of low-energy explosions of RSGs in the context of failed SNe. \citet{2013ApJ...768L..14P} estimated SBO characteristics semi-analytically, in both the radiation-pressure-dominated and gas-pressure-dominated regimes, assuming the surface density profile followed an $n = 3/2$ polytrope.  They assume that the total energy budget is the internal energy, $E_i$, contained in the mass layer between the surface of the ejecta and the $\tau = c / v_{\rm sh}$ surface; duration is assumed to be the diffusion time through that layer. The peak luminosity is then estimated as $E_i$ over the diffusion time. In a way, this gives the average luminosity over the SBO peak (not the peak luminosity that we measure).  The formulae of \citealt{2013ApJ...768L..14P} over estimate the luminosity of our models by factors of $1.5 - 9$ and differ in duration by factors of $2-8$ (the difference in duration is an over-prediction in some cases and an under-prediction in others).  Some of this disagreement may come from the fact that our stellar models do not resemble an $n = 3/2$ polytrope at the surface. However, as we noted above, the surface structure is sensitive to our choices made in the stellar modeling. 

More recently, \citet{Lovegrove2017} simulated low-energy explosions of two RSGs (with initial masses of $15$ and $25 \msun$). Their explosions ranged in energy from $10^{47}$ to $10^{51}$ ergs.  The focus of their paper was to study SBO in detail using multigroup, flux-limited-diffusion, radiation-hydrodynamical simulations with the \texttt{Castro} code, but they also used the \texttt{Kepler} stellar evolution code to simulate the full light curves of explosions of their $15\msun$ RSG. Their models were also in 1D, but they corrected for light travel time effects and their multigroup opacities allowed for non-thermal photon distributions. Our model with a ZAMS mass of $14\msun$ ($\mej \approx 7.43\msun$ and $R_\star = 1089 \rsun$) is the closest to their $15\msun$ model ($\mej \approx 8.52\msun$ and $R_\star \approx 862\rsun$).  Comparing our models 14E48, 14E49, and 14E50 to models B15, C15, and E15, respectively, from their fig. 18 (and noting that their time axis is log-scaled), our lightcurves are very similar and there is excellent agreement between duration and luminosity on the plateau.  Their B15, C15, and E15 models have $\tpl$ of $\sim$ 300, 190, and 125-130 days, while our models of similar energies have, respectively, $\tpl =$ 315, 188, and 130 days.  The luminosities on the plateau for their models B15, C15, and E15 are $\approx$ $1-2$, $6-10$, and $50-80$ $\times 10^{40}$ ergs s$^{-1}$, respectively, our models of similar energies, respectively, have $\lpl =$ 1, 12, and 78  $\times 10^{40}$ ergs s$^{-1}$.   There is poorer agreement with our measured SBO peaks and SBO durations and theirs.  After correcting for the difference in radius (\citealt{2013ApJ...768L..14P} gives $t_{\rm sbo} \propto R_\star^{2.19}$), our SBO durations differ by a factor of $1.5-2$ and $L_{\rm peak}$ in our models is a factor of $2-10$ dimmer than theirs (with the largest disagreement at $10^{50}$ erg). Although $L_{\rm peak}$ and $t_{\rm sbo}$ differ, our models agree in the radiated energy, $L_{\rm peak}\times t_{\rm sbo}$, which is expected to be less sensitive to the stellar structure. Thus, the difference in duration and brightness of the shock breakout signature are reasonable given the very different treatments of radiation and the differences in density profiles at the stellar surface. 

\citet{2010MNRAS.405.2113D} performed a suite of simulations in which they exploded a 11$\msun$ RSG (final mass of 10.6$\msun$ and final radius of 587$\rsun$) varying the energies from those below the binding energy of the H/He layers to energies much larger than the binding energy. They also varied where they deposited the bomb and the duration of the energy deposition. Most similar to our simulations are their models s11\_1\_w3 and s11\_3\_w3. The energy was deposited nearly instantaneously near the base of the hydrogen envelope and sufficient energy was deposited to robustly unbind the hydrogen envelope. Model s11\_1\_w3 ejected $7.6\msun$ of material with a final kinetic energy of $1 \times 10^{49}$ erg. The lightcurve for this model is the blue line in the left panel of their fig.~10.  From the plot, the luminosity at the mid-point of the plateau (where we measure $\lpl$ in our models) is $\approx 1\times 10^7 \lsun$ and the duration (from their table 2) is 181 days.  Model s11\_3\_w3 also ejected $7.6\msun$ of material but with a final kinetic energy of $4.1 \times 10^{49}$ erg. The lightcurve is plotted with a blue line in the right panel of their fig.~10.  The luminosity at the mid-point of the plateau is $\approx 3\times 10^7 \lsun$ and $\tpl$ =  142 days.  The shapes of their lightcurves are very similar to our RSG explosions, where the luminosity rises monotonically from a local minimum at the end of the shock cooling feature to a maximum just before the end of the plateau. To quantitatively compare our models to theirs, we use eqs.~\eqref{eq:lpl_fit} and ~\eqref{eq:tpl_fit} to predict $\lpl$ and $\tpl$ for their models.  Their actual $\lpl$ values are a factor of $1.3$ higher than our prediction; $\tpl$ agrees within a few percent.

Finally, \citet{tsuna_transients_2025} simulated $10^{48}$ and $10^{49}$ erg explosions of two RSG models in the context of failed SNe.  They ran models with and without a dense CSM using the supernova explosion code \texttt{SNEC}.  They fed the explosions for a duration of $\sim$ 1 day at a radius of $\sim$10 $\rsun$.  Applying our numerical fits to their models without CSM, we find good agreement with $\lpl$ but their event durations are significantly longer ($\sim 30\%$) than predicted from our eq.~\eqref{eq:tpl_fit}. One potential difference could be the inclusion of hydrogen and helium recombination in the \texttt{SNEC} equation of state, which is not included in our models. Recombination energy becomes more important as the ejecta becomes less radiation-pressure dominated, as we discuss in Sec.~\ref{sec:recombination} below.  On the other hand, the calculations of \citet{2010MNRAS.405.2113D} employed a very sophisticated EOS that included all of the atomic excitations/ionizations/recombinations; our results agree with the durations of their models within a few percent.  Thus, the disagreement in $\tpl$ between our models and \citet{tsuna_transients_2025} could be driven by a variety of factors. 

\subsection{Limitations of our models}
\subsubsection{The importance of recombination photons in low energy explosions}
\label{sec:recombination}
When the recombination front sweeps inward through the expanding ejecta, the gas outside the recombination front emits recombination photons.  If the material is very optically thin, these photons will free-stream to infinity and add to the luminosity of the light curve.  In regions of higher optical depth, these photons can be absorbed by the material and heat the gas.  Although our opacities do account for the ionization state of the gas, we do not capture the effect of recombination photons on our lightcurves or on the thermal state of the gas. In SNe IIp, these effect are not important because (1) the total energy radiated in photons is much larger than the total recombination energy budget of the ejecta and (2) the emitting region is very radiation-pressure dominated. The low-energy explosions that we consider here are less radiation-pressure dominated, thus the recombination photons could play an important role \citep[e.g., ][]{2022ApJ...938....5M}. 

To check the importance of recombination energy/photons for our models, we estimate the total radiated luminosity as $\lpl \times \tpl$ and the total recombination energy budget as $E_{\rm rec} \approx [(X_{\rm H} \mej / m_p) \times 13.6{\rm eV}] +  [(X_{\rm He} \mej / 4 m_p) \times 10.2{\rm eV}]$, where $X_{\rm H}$ and $X_{\rm He}$ are the hydrogen and helium mass fractions of the ejecta, respectively. The final column of Table~\ref{tab:athmodels} shows the ratio $\lpl\tpl/E_{\rm rec}$.  For our H-rich, $10^{48}$ explosions, (12E48, 14E48, and 18E48), this ratio is $1.6 - 2.5$.  Thus, for these models, including hydrogen recombination in the equation of state could impact the lightcurve.

Finally, we note that we also simulated a set of H-rich explosions with $\extra = 10^{47}$ erg. For those models, $E_{\rm rec}> \lpl\tpl$ and the emitting region on the plateau was very gas-pressure dominated.  We omit those calculations from this work as including hydrogen recombination in the equation of state would likely significantly change the lightcurve. Such calculations will be the subject of future work.

\subsubsection{Instantaneous energy deposition and location of the deposited energy}
For this initial study, our simulations drive explosions by initializing the stellar profile with an over-pressure region at the base of the convective hydrogen envelope.  By depositing the energy instantaneously, we are assuming the energy-deposition time is much shorter than the dynamical time of the envelope.  In nature, weak, hydrogen-rich transients with RSG and YSG progenitors may be the result of eruptive mass loss, common envelope events, sub-energetic neutrino-driven SN, and failed SN.  In the case of common envelope events, stellar mergers, and failed SN, the energy my be injected over timescales approaching the dynamical time of the envelope (or longer).  For example, for the failed SN mechanism studied by \citet{paperII}, in which accretion of a small fraction of the convective envelope onto the newborn BH drives an explosion of the hydrogen envelope, the outflow will continue to be fed with mass and energy until accretion totally shuts off. This may take a full dynamical time of the envelope ($\sim$ months in the case of RSGs).  In the case of stellar mergers / common envelope events, the companion spirals into the donor envelope on an envelope dynamical time, depositing energy over an extended time and also over an extended range of radii within the donor envelope. 

\citet{2010MNRAS.405.2113D} explored the effects of varying the location of energy-deposition and the timescale over which the energy is deposited. Their fig.~11 shows light curves for $4\times10^{49}$ erg explosions of their fiducial $10.6\msun$ star with energy deposition times of 10 seconds, 1 hour, 1 day, 1 week, and 1 month (the dynamical time of the star is $\sim 80$ days).  The lightcurves are nearly identical for deposition times up to a week. However, depositing the energy over a month, which is a significant fraction of the $\sim 80$-day dynamical time of the star, extends the duration from 140 days to 170 days. They also varied the mass coordinate of the deposited energy. In their models s11\_01, s11\_1, and s11\_2, the deposited energy is less than the binding energy of the mass exterior to the radius of energy deposition. The light curves for those models are shown in their fig.~10 in purple, dark blue and medium blue, respectively.  For those explosions, the duration of the lightcurve is longer than our eq.~\eqref{eq:tpl_fit} predicts.  One reason for this could be that the final kinetic energy of the ejecta reported in their table 1 is not the asymptotic kinetic energy of the unbound ejecta (much of the material is still bound).  In addition, we found in some of our preliminary models that the presence of a marginally-bound layer added an extension to the lightcurve (e.g. the extra ``step'' in the medium blue light curve for model s11\_2 in their fig.~10) as the photosphere recedes through this bound fallback. Thus, explosions that probe the transition between bound and unbound have different properties than those that robustly unbind the envelope. 

\subsubsection{Stellar surface and circumstellar environment}
Here we have simulated stars in 1D with a simple exponential profile in density beyond the surface of the stellar model.   
In nature, the surfaces and circumstellar environments of RSGs and YSGs are complex.  In 3D models, radiation hydrodynamic simulations of partially-stripped YSGs show extended structures of bound material that are lifted from the stellar surface and fall back \citep{2025arXiv250812486G}. For RSGs, dense circumstellar material has been seen in early time observations of many SN IIp \citep[see, e.g., and references within][]{2024ApJ...970..189J} and may be present for the vast majority of RSGs at the time of explosion \citep{2022MNRAS.517.1483D}.  This dense CSM could be the result of a wind \citep{2017ApJ...838...28M}, pre-supernova mass loss \citep{Wu_2021}, or an extended chromosphere that is bound to the star \citep{2024OJAp....7E..47F}.  The structures outside of the stellar surface impact the shock breakout signatures of these explosions. Motivated by this, \citet{tsuna_transients_2025} studied the effect of this chromosphere on low energy explosions of two RSGs like those studied here.  Although the lightcurve plateau was unaffected by the CSM, the chromosphere extended shock breakout from $\sim$1 day to 3-10 days. As we have noted above, the 3D convective structure at the surface of the star blurs the shock breakout by a similar amount.

\subsection{Failed supernova candidates}
Failed SNe are a large motivation for this work so we discuss them in detail in this section and consider the implications of our models for the failed SN candidates that have been reported in the literature. To date, there are two observed transients that are strong candidates for failed SNe.\footnote{\citet{2017MNRAS.467.3299K} have proposed an alternative scenario, in which mass-loss in a binary system can temporarily obscure the system by dust in the orbital plane and can redirect some of the luminosity away from the observer viewing the system through this geometrically thick obscuring medium (though also see the radiative transfer calculations of \citealt{2023arXiv230511936K} which counter this claim). Here we consider the events in the context of low-energy explosions of single massive stars.}  The first was identified in the galaxy NGC 6946 as part of a monitoring campaign of RSGs in nearby galaxies using the Large Binocular Telescope \citep[LBT; ][]{2008ApJ...684.1336K,2015MNRAS.450.3289G}.  In this event (hereafter, N6946-BH1), a massive supergiant showed an optical brightening that began in 2014 and lasted $3-11$ months\footnote{The uncertainty in the duration is due to the $3-4$ month cadence of the LBT observations.} before dimming by 5 magnitudes in the optical. Coincident with the effective disappearance of the star in the optical, the source shows a power-law fall off in the infrared \citep{2015MNRAS.450.3289G,2017MNRAS.468.4968A,2021MNRAS.508.1156B,2021MNRAS.508..516N}.  Followup observations in 2024 with JWST confirm that the bolometric luminosity is still $\approx0.1-0.15$ times the progenitor luminosity \citep{2024ApJ...962..145K,2024ApJ...964..171B}.  The progenitor was originally classified as a $\sim25\msun$ RSG \citep{2017MNRAS.468.4968A} but \citet{2019RNAAS...3..164H} later argued that the progenitor is hotter than typical RSGs and, thus, could be a YSG (more on that below).  

The second strong candidate for a failed SN (hereafter, M31-2014-DS1) was identified by \citet{2024arXiv241014778D} in the Andromeda galaxy (Messier 31) as part of a search for mid-infrared (MIR) transients in archival NEOWISE data. The source underwent a $~50\%$ brightening in MIR flux from 2014 to late 2017 with a corresponding dimming in the optical.  Since late 2017, the source is undetected in the optical and the MIR flux has steadily faded to below the progenitor flux. During the $\sim1000$ day MIR brightening, the bolometric luminosity was roughly constant, before declining as a power-law over the next $\sim1000$ days. The most recent observations show that the source is a factor of 10 dimmer in total light compared to the progenitor star and at least a factor of $10^4$ dimmer in visible light (i.e. down to the upper limits in the optical as the source is undetected at those wavelengths). Using archival HST and Spitzer data from the UV to MIR bands, they are able to fit the progenitor SED with a $\log(L/L_\odot) \approx 5$, $T_{\rm eff}\approx 4500$ K YSG surrounded by a dust shell with effective temperature $\approx 800$ K. Although stellar mergers and eruptive mass loss can cause substantial optical dimming of stars, the bolometric luminosity should not dim, in general.  The reduction of the bolometric luminosity to $10\%$ of the progenitor luminosity in the M31-2014-DS1 source suggests the cessation of a nuclear burning source and likely death of the star. Additionally, the $t^{-5/3}$ decline in bolometric luminosity after a $\sim$1000-day constant bolometric luminosity is suggestive of accretion onto a nascent BH.   

In M31-2014-DS1, there was no optical outburst detected within the cadence of the available optical data, implying the optical outburst was very dim or very short (less than $\sim$80 day).  \citealt{2024arXiv241014778D} use these to place limits on the mass and energy of any ejecta in the outburst/explosion. Additionally, if the $t^{-5/3}$ decline in bolometric luminosity is related to the expected $t^{-5/3}$ accretion of weakly bound fallback material following a  $\sim$spherical low-energy explosion, then the accretion rate must be below Eddington.  The transition from constant bolometric luminosity to the power-law decline at $\sim1000$ days then also constrains the time-dependent accretion rate onto the BH.  Any model must self-consistently fit the progenitor properties, the range of possible ejecta properties for a range of explosion energies of that progenitor, and the properties of the fall back that would be generated in such an explosion. \citealt{2024arXiv241014778D} find the available data to be most consistent with a $\sim$$10^{47} - 10^{49}$ erg explosion of a $\sim$5.0 $\msun$ YSG with a low-mass hydrogen envelope of $\sim0.28\msun$. For that range of energies, the explosion unbinds $\sim0.2$ $\msun$ of the envelope. This implies a compact-object mass of $\sim 4.8\msun$, thus the remnant is a BH, not a neutron star.   

\citealt{2024arXiv241014778D} applied a similar method to reanalyze N6946-BH1. They show that a hydrogen-rich RSG is not consistent with the data. Using a hotter effective temperature of 4560 K \citep[argued for by ][]{2019RNAAS...3..164H}, they match the progenitor with a $\sim$7$\msun$ YSG with an envelope mass of 0.6$\msun$. They constrain the ejecta mass and explosion energy using a similar procedure as for the M31 event (but also including the constraints placed by the observed optical outburst in this case). They find that the transient is consistent with a failed SN of the YSG progenitor that led to a $\sim$$10^{47} - 10^{49}$ erg explosion which ejected $\sim0.3$ $\msun$ of the envelope.

\begin{figure}
\centering
\includegraphics[width=\columnwidth]{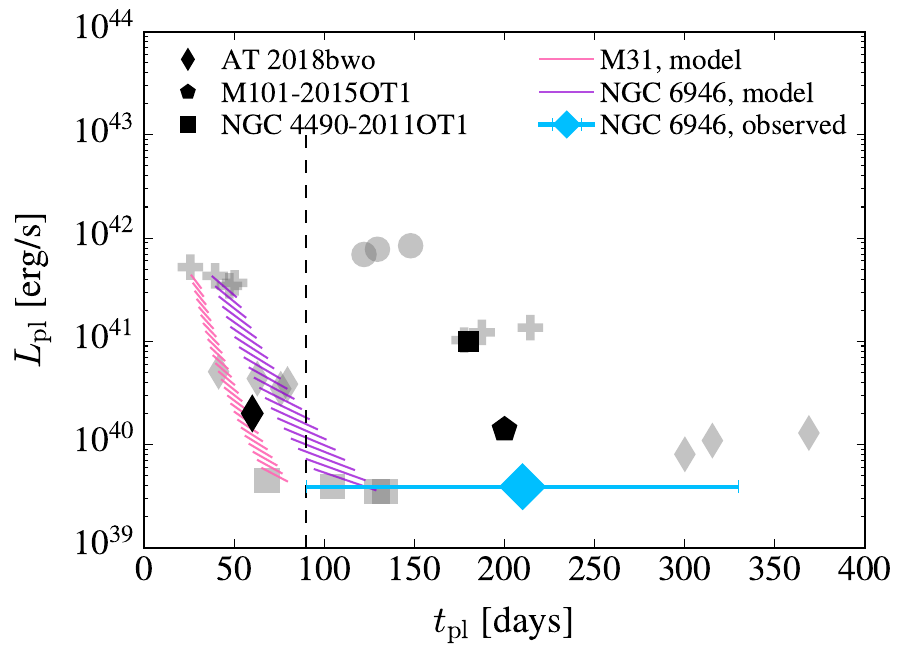}
\caption{Pink and purple hatched regions: plateau luminosity and duration expected for the two failed SN candidates proposed by \citealt{2024arXiv241014778D} based on our fitting formulae for $\lpl$ and $\tpl$. The hatched regions cover a range in ejecta mass (from that estimated by \citet{2024arXiv241014778D} to the total mass of the hydrogen envelope) for the range of ejecta kinetic energies expected for failed SN for the two proposed progenitors ($10^{47} - 10^{49}$ erg).  The vertical dashed line is the maximum duration of an outburst for M31-2014-DS1 that could have been missed in the optical.  From our scalings, the estimated progenitor and explosion parameters for M31-2014-DS1 proposed by \citealt{2024arXiv241014778D} fit within the observational constraints.
The blue diamond and error bar is the observed optical transient for N6946-BH1 \citep{2017MNRAS.468.4968A}.   For this event, our scalings suggest that the ejecta mass was at the higher end of the mass range (the entire H envelope, $\sim0.6\msun$) and at the low end of the energies considered. Black symbols are LRNe with supergiant progenitors. Grey symbols in the background are the same as in the right panel Fig.~\ref{fig:L_t_peaks}, showing  $\lpl$ and $\tpl$ for our suite of simulations.  Failed SN and LRNe occupy very similar parameter space. Lightcurve modeling of these transients is essential for differentiating them photometrically.}
\label{fig:FSN}
\end{figure}

In Fig.~\ref{fig:FSN}, we consider the luminosity and duration of the transients that could result from ejection of the hydrogen envelopes of the two failed SN progenitors proposed by \citealt{2024arXiv241014778D}. Their modeling of the explosion and fallback accretion is approximate so we consider a somewhat wider range of ejecta masses than they estimate from their fallback modeling.  For M31-2014-DS1, we take $0.2 \le \mej/\msun \le 0.28$ while for N6946-BH1 we consider $0.3 \le \mej /\msun \le 0.6$. Here, the lower limits are the lower estimates from \citealt{2024arXiv241014778D} while the upper limits are the total masses of the hydrogen envelopes of the two progenitors.  We use our fitting formulae of eqs.~\eqref{eq:lpl_fit} and \eqref{eq:tpl_fit} to estimate the plateau luminosity and duration of $\eej = $$10^{47} - 10^{49}$ erg explosions with those ejecta masses.  The resulting ranges in $\lpl$ and $\tpl$ are shown as hatched regions in Fig.~\ref{fig:FSN}. The pink region corresponds to M31-2014-DS1.  For that event, there was no optical outburst detected so the optical outburst must have been shorter than $\sim80$ days (the vertical dashed line in Fig.~\ref{fig:FSN}). The entire pink region falls to the left of that line, thus all of the ejecta masses and energies we plot are acceptable within the observations.  The purple region shows the expected $\lpl$ and $\tpl$ for N6946-BH1. We also plot the observed optical outburst with the blue diamond in Fig.~\ref{fig:FSN}. The error bar reflects that they cannot constrain the duration to better then $3-11$ months. We note that the luminosity of the outburst could be a lower-limit since the transient could have been brighter on shorter timescales than they were sensitive to.  The purple region to the left of $\tpl\approx 3$ months is not permitted as the duration would be too short.  Thus, the lower $\eej$ and larger $\mej$ events are more favored (the lower right section of the purple region). The purple region meets the blue error bar at $\mej \approx 0.6\msun$ (the entire hydrogen envelope of the YSG model) and $\eej \approx 10^{47}$ erg.  

By adopting eqs.~\eqref{eq:lpl_fit} and \eqref{eq:tpl_fit}, the pink and purple regions assume the explosion is driven over much shorter timescales than the dynamical time of the envelope.  This is a reasonable assumption if the dominant mechanism for unbinding the mass is the neutrino-mass-loss mechanism. If the dominant driver of the explosion is, instead, the angular momentum barrier faced by the infalling convective layer, then the energy is fed on longer timescales that could increase the duration of the event. This mechanism is probably required to explain the luminosity of the fallback accretion as the spherically-symmetric explosions and fallback of the neutrino-mass-loss mechanism alone would produce fallback luminosities that are far dimmer than those seen in the two failed SN candidates \citep{2025arXiv250716893F}.  

More careful modeling of the failed SN explosion mechanisms and fallback of the marginally bound material would be needed to better constrain $\mej$ and $\eej$ for M31-2014-DS1. For N6946-BH1, the large uncertainty in observed duration of the optical outburst would likely limit the ability of such models to constrain $\mej$ and $\eej$ further, though continued monitoring of the late-time light curve in the infrared could permit fruitful comparisons to more detailed models of the fallback.

\section{Summary and Conclusions}
\label{sec:summary} 

A variety of phenomena can lead to a low-energy ejection of the material from the convective hydrogen envelope of an evolved supergiant.  These include common envelope events / stellar mergers in binary systems, under-energetic core collapse SNe, electron-capture SNe, eruptive mass loss, and failed SNe.  Observationally, these events may appear as luminous red novae, sub-luminous SN IIp, intermediate luminosity red transients, or as stellar brightenings that are followed by the disappearance of the star.

We have used one-dimensional, radiation hydrodynamic simulations to model the simplest case of ejection of the hydrostatic envelope of RSGs and partially-stripped YSGs following instantaneous injection of energy near the base of the convective hydrogen layer. Our RSG models simulate explosions with energies of $10^{48}, 10^{49},$ and $10^{50}$ ergs and with ejecta masses of $\approx 7 - 9 \msun$. Our YSG explosions have energies of $10^{47}, 10^{48},$ and $10^{49}$ ergs and range in ejecta mass from $\approx 0.2 -0.6 \msun$. 

The following summarizes our main results:

\begin{enumerate}
    \item The explosions have ejecta velocities of $\sim$few hundred to $\sim$few thousand km s$^{-1}$. Owing to their lower envelope masses, explosions of stripped YSGs have higher ejecta velocities than explosions of RSGs with the same energy. 
    \item The lightcurves are similar to those of SNe IIp with an initial peak due to shock breakout / shock cooling followed by a long plateau mediated by hydrogen recombination (Figs.~\ref{fig:lightcurves_hrich} and \ref{fig:lightcurves_hpoor}). There is no nucleosynthesis in low energy explosions of the hydrogen envelope so there is no radioactive tail in the lightcurve; the lightcurve ends at the end of the plateau, where, in the case of failed SN, it will transition to emission by fallback accretion onto the BH \citep{2025arXiv250716893F}.
    \item The peak luminosity ranges from $\sim$few $\times 10^{40}$ to  $\sim$few $\times 10^{43}$ erg s$^{-1}$ and the duration of the shock breakout / shock cooling peak ranges from 7 to 58 hours (Fig.~\ref{fig:L_t_peaks}, left panel, and Table~\ref{tab:athmodels}).
    \item Luminosities on the plateau are $\sim$few $\times 10^{39}$ to $\sim10^{42}$ erg s$^{-1}$. The hydrogen-rich, RSG explosions have durations of $100-400$ days while hydrogen-poor, YSG explosions have overall durations of 26 to 150 days (Fig.~\ref{fig:L_t_peaks}, right panel, and Table~\ref{tab:athmodels}).
    \item The scalings of \citet{Popov1993} for the plateau luminosity ($\lpl$) and lightcurve duration ($\tpl$) capture reasonably well our RSG (YSG) models with explosions energies $\gtrsim 10^{49}$ erg ($\gtrsim10^{48}$ erg). We provide more accurate fitting formulae for $\lpl$ and $\tpl$ in eqs.~\eqref{eq:lpl_fit} and \eqref{eq:tpl_fit}, respectively.
    \item Our RSG models with explosion energies $\eej \gtrsim 10^{49}$ are radiation-pressure dominated on the plateau; at $\eej\sim10^{48}$ erg, the emitting region has $P_{\rm rad} \sim P_{\rm gas}$ and absorption opacities become important at the photosphere (Fig.~\ref{fig:emit_layer_12}). Our YSG models, on the other hand, are radiation-pressure-dominated on the plateau for $\eej \gtrsim 10^{48}$ and gas pressure becomes important at the photosphere near $\eej \sim 10^{47}$ erg (Fig.~\ref{fig:emit_layer_12.5}).  
    \item Using our fitting formulae for $\lpl$ and $\tpl$, we estimate the outbursts that could accompany the failed SNe candidates N6946-BH1 and M31-2015-DS1 using the estimated explosion parameters of \citealt{2024arXiv241014778D}. We find that their values are consistent with the observations, though our calculations predict the explosion energy (ejecta mass) should be at the lower end (higher end) of their estimates for N6946-BH1 (Fig.~\ref{fig:FSN}).
\end{enumerate}

Failed SNe, luminous red novae, and other low-energy mass ejections from RSGs and YSGs are observationally quite similar.  Fig.~\ref{fig:FSN} shows that the low-energy explosions studied in this paper, including failed SN, have similar luminosities and durations to extra-Galactic luminous red novae.  Though not shown, these transients are characterized by an early blue peak followed by a red plateau and have ejecta velocities of $\sim$100s to $\sim$ few thousand km s$^{-1}$.   

In the optical, the Vera C. Rubin Observatory's Legacy Survey of Space and Time \citep[LSST; ][]{2019ApJ...873..111I} will find $300-1500$ luminous red novae per year \citep[peak in the r-band of $M < -13.5$ mag]{2023ApJ...948..137K}.  To estimate the number of failed SNe that LSST is sensitive to, we construct broadband light curves for a sample of explosions in Appendix~\ref{sec:appendix}.  For RSGs, failed SNe have explosions energies of $10^{48}$ to $10^{49}$ erg. Models 12E48 and 12E49 have peak absolute AB magnitudes of $M_r =-13.5$ and $-14.8$ mag in the LSST $r$ band, respectively (Fig.~\ref{fig:broadband_12M_hrich}).  Taking $M_r \approx -14$ mag and assuming a sensitivity of $m_r\approx 24.2$ mag, LSST is sensitive to events within a volume $V_r \approx (220 {\rm \, Mpc})^3$ (where we have divided by 2 given Rubin observes only half of the sky).  

Based on 11 years of data monitoring 27 nearby galaxies for failed SN with the LBT,  \citet{2021MNRAS.508..516N} estimate a fractional failed SN rate of $f_{\rm FSN} = 0.162^{+0.232}_{-0.125}$  relative to the core-collapse SN (CCSN) rate (this assumes one failed SN, N6946-BH1, relative to the 8 CCSNe found in their survey to date).  The Sloan Digital Sky Survey II Supernova Survey finds a CCSN rate of $10.6^{+1.9}_{-1.9} \times 10^4 (h/0.7)^3$ Gpc$^{-3}$ yr$^{-1}$ \citep{2014ApJ...792..135T}, while the Zwicky Transient Facility Bright Transient Survey finds a rate of $10.1^{+5.0}_{-3.5} \times 10^4$ Gpc$^{-3}$ yr$^{-1}$ \citep{2020ApJ...904...35P}.  Adopting an average CCSN rate of $R_{\rm CCSN} = 10.35 \times 10^4$ Gpc$^{-3}$ yr$^{-1}$, then the failed SN rate is roughly $f_{\rm FSN} R_{\rm CCSN} \approx 1.64\times 10^4$ Gpc$^{-3}$ yr$^{-1}$.   Thus, there would be $\sim700$ failed SN per year in the volume probed by LSST.  

The above estimates assume that the shock breakout and cooling peaks are as bright as in our 1D models.  We noted in Sec.~\ref{sec:sbo} that 3D effects at the RSG and YSG surface will impact the SBO signature.  The convective structure spreads out the SBO signature from a duration of $t_{\rm sbo}$ to a duration of $t_{\rm cross} \sim\Delta R/v_{\rm sh}$. The energy radiated during SBO in our models can be estimated as $E_{\rm sbo} \sim L_{\rm peak}\times t_{\rm sbo}$. Accounting for the 3D structure, this energy would instead be emitted over a time $t_{\rm cross}$, so the luminosity accounting for the 3D effects can be estimated as $L_{\rm 3D} \approx E_{\rm sbo}/t_{\rm cross}$.  Then $L_{\rm 3D} / L_{\rm peak} \approx v_{\rm sh}t_{\rm sbo}/\Delta R$. For $\Delta R = 100\rsun$ (Sec.~\ref{sec:sbo}) and taking $v_{\rm sh} \approx 342$ km s$^{-1}$ and $t_{\rm sbo} = 24$ hours (for model 12E48 in Table.~\ref{tab:athmodels}), $L_{\rm 3D} / L_{\rm peak} \approx 0.5$, which lowers the absolute magitude by $\approx 0.93$ mag. This reduces the failed SN rate in the volume probed by LSST to $\sim 200$ yr$^{-1}$. We note that our estimated rates do not fold in the LSST survey cadence of $\sim$few days. The SBO feature is less than a day, so a fraction of SBO peaks would be missed. At the same time, the plateau is nearly as bright (or brighter) than the SBO feature in the $r$ band for most of the models (especially the H-poor models and models with $\extra \gtrsim 10^{49}$ ergs). The plateau lasts 10s to 100s of days, so would be sampled repeatedly in the few-day cadence of LSST. 

The vast majority of LSST alerts will not be followed up spectroscopically; careful light curve calculations such as those presented in this work will be important for characterizing these transients. In addition, the low-energy hydrogen envelope ejections that are the focus of this work rapidly form dust, obscuring the system in the optical \citep{2025arXiv250803932K}. Indeed, the failed SN candidate in M31 was missed in the optical but discovered as a MIR transient with WISE.  We are entering an exciting time for IR transients with the Wide-field Infrared Transient Explorer \citep[WINTER,][]{2020SPIE11447E..9KL,2020SPIE11447E..67F} now performing a wide-field photometric survey. WINTER will soon be joined by the Dynamic REd All-sky Monitoring Survey \citep[DREAMS, ][]{2020SPIE11203E..07S} in the south. Complementary to all of these photometric surveys is the recently launched Spectro-Photometer for the History of the Universe, Epoch of Reionization and Ices Explorer \citep[SPHEREx, ][]{2018arXiv180505489D}, which will perform an all-sky spectroscopic survey in the optical and near-IR.   We are, of course, in the age of JWST, which extends our reach to the far IR, enabling access to the full SED of the post-outburst remnants of these systems. Such access to the complete SED is important for differentiating between stellar mergers, BH-births in failed SN, and other dusty outbursts from massive stars.

\section{Software and third party data repository citations} \label{sec:cite}






The \mesa inlists used to generate out progenitor models will be made available on Zenodo with the final manuscript. 

\begin{acknowledgments}
The work presented in this paper was conducted in xu\v{c}yun (Huichin), the native homelands of the Ohlone people on which Berkeley sits and in Lenapehoking, the native homelands of the Lenape peoples that include  Princeton and New York City. 

The authors wish to thank Kishalay De, Tamar Faran, Jared A. Goldberg, Morgan Macleod, Sophie L. Schr\o der, and Benny T.-H. Tsang for useful conversations. This work benefited from interactions with Lars Bildsten, Bill Paxton, Jim Fuller, and Daichi Tsuna that were funded by the Gordon and Betty Moore Foundation through Grant GBMF5076.

A.A. gratefully acknowledges support from the U.C. Dissertation-Year Fellowship, the Maria Cranor Fellowship, the U.C. President's Postdoctoral Fellowship Program, the National Science Foundation Graduate Research Fellowship under Grant No. DGE 1752814, the Gordon and Betty Moore Foundation through Grant GBMF5076 and the Flatiron Institute.  E.Q. was supported in part by a Simons Investigator award through the Simons Foundation. The Flatiron Institute is supported by the Simons Foundation.

The simulations presented were performed on the Rusty and Popeye clusters at the Flatiron Institute.  Preliminary calculations made use of computational resources managed and supported by Princeton Research Computing, a consortium of groups including the Princeton Institute for Computational Science and Engineering and the Office of Information Technology's High Performance Computing Center and Visualization Laboratory at Princeton University. 
\end{acknowledgments}

\begin{contribution}

\end{contribution}

%

\software{This project was made possible by the following publicly available software: \texttt{Astropy}\footnote{http://www.astropy.org} \citep{astropy:2013, astropy:2018}, \ath \citep{2020ApJS..249....4S}, \texttt{Matplotlib} \citep{Hunter:2007}, \mesa \citep{2011ApJS..192....3P,2013ApJS..208....4P,2015ApJS..220...15P,2018ApJS..234...34P,2019ApJS..243...10P,2023ApJS..265...15J},  \texttt{MESA SDK} \citep{richard_townsend_2020_3706650}, \texttt{NumPy} \citep{harris2020array}, \texttt{yt}\footnote{https://github.com/yt-project/yt} \citep{2011ApJS..192....9T} }


\appendix
\section{Broadband Lightcurves}
\label{sec:appendix}
In this section, we compute approximate broadband lightcurves for a subset of our models in the LSST $r$, $g$, and $u$ bands by post-processing our frequency-independent simulations under the assumption that the photosphere emits as a blackbody with temperature $\teff$ (this assumption will be replaced with multigroup calculations in future work).  We obtained the transmission profiles, $S(\nu)$, for LSST from the SVO Filter Profile Service\footnote{\href{https://svo2.cab.inta-csic.es/theory/fps/fps.php}{https://svo2.cab.inta-csic.es/theory/fps/fps.php}} \citep{2020sea..confE.182R, 2012ivoa.rept.1015R, 2024A&A...689A..93R}.  

\begin{figure*}
\centering
\includegraphics[width=0.8\textwidth]{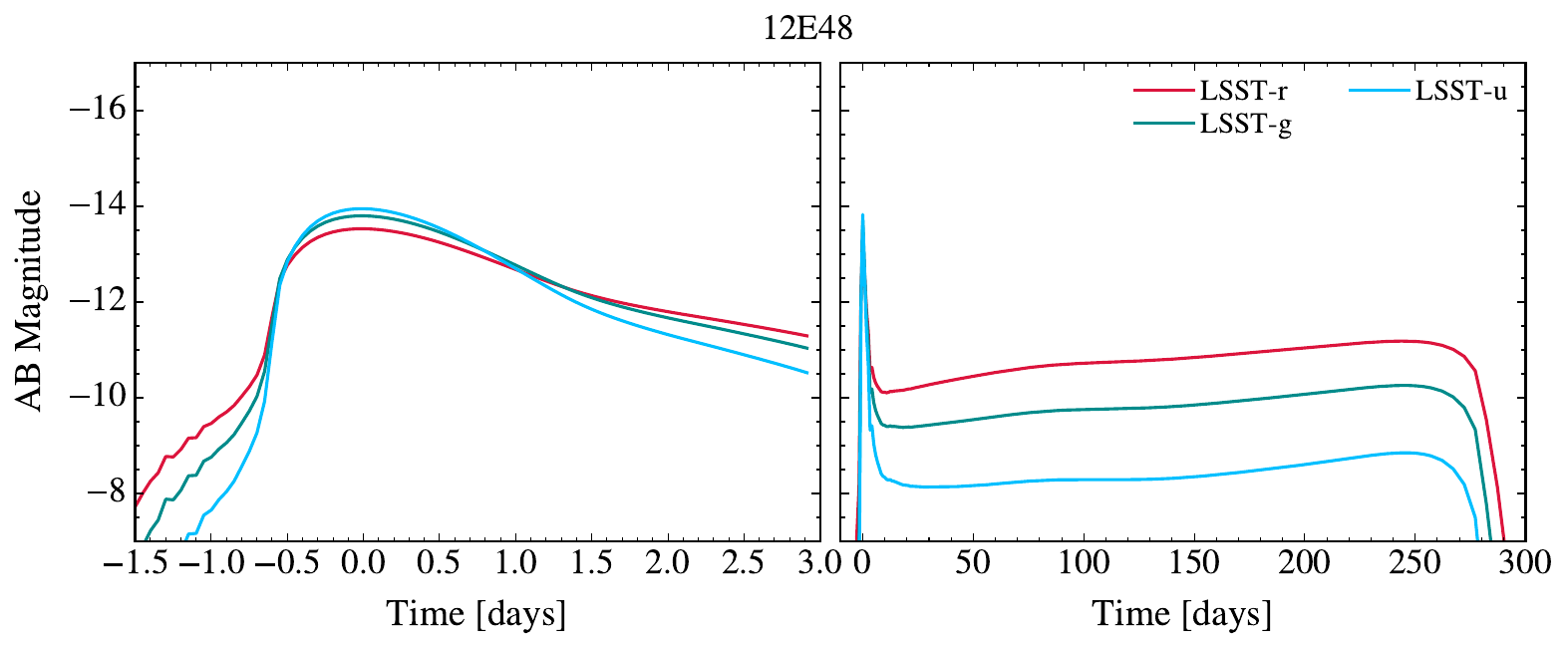}
\includegraphics[width=0.8\textwidth]{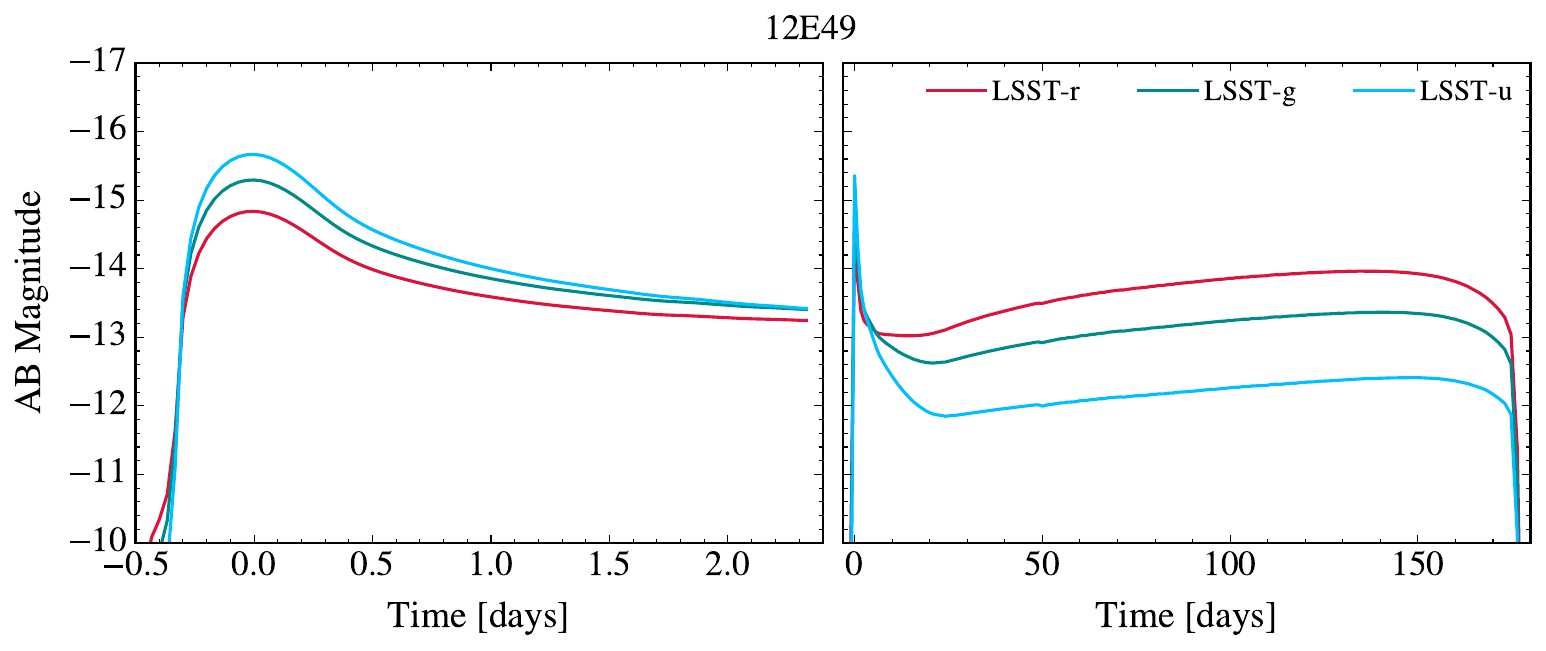}
\includegraphics[width=0.8\textwidth]{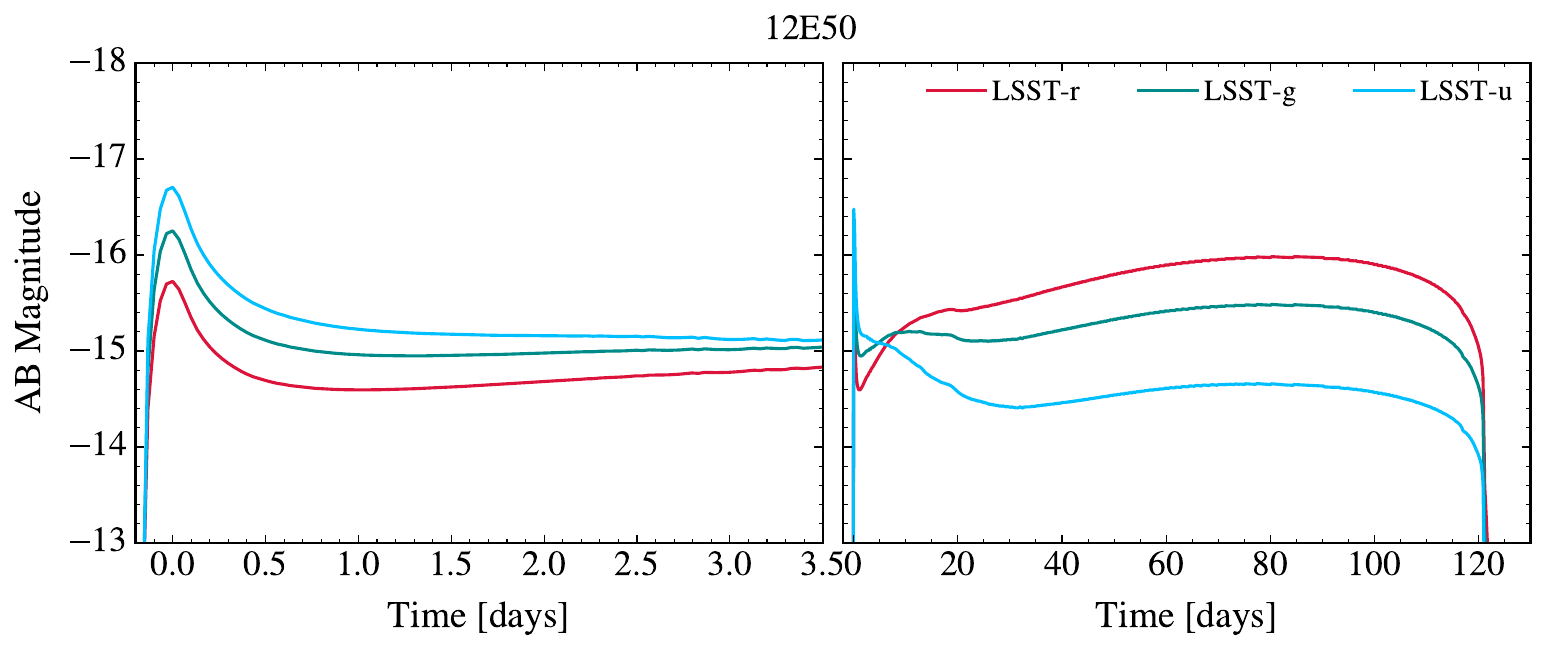}
\caption{Absolute (AB) magnitude versus time in the LSST $r$, $g$, and $u$ bands for explosions of our RSG model with $\mzams = 12\msun$. In each row, the left panel is zoomed in to highlight the shock breakout and cooling feature while the right panel shows the full lightcurve. The top, middle, and bottom rows are for $\log(\extra /{\rm erg})$ = 48, 49, and 50, respectively. Note that the $x-$ and $y-$ axes limits are different in each row.}
\label{fig:broadband_12M_hrich}
\end{figure*}

\begin{figure*}
\centering
\includegraphics[width=0.8\textwidth]{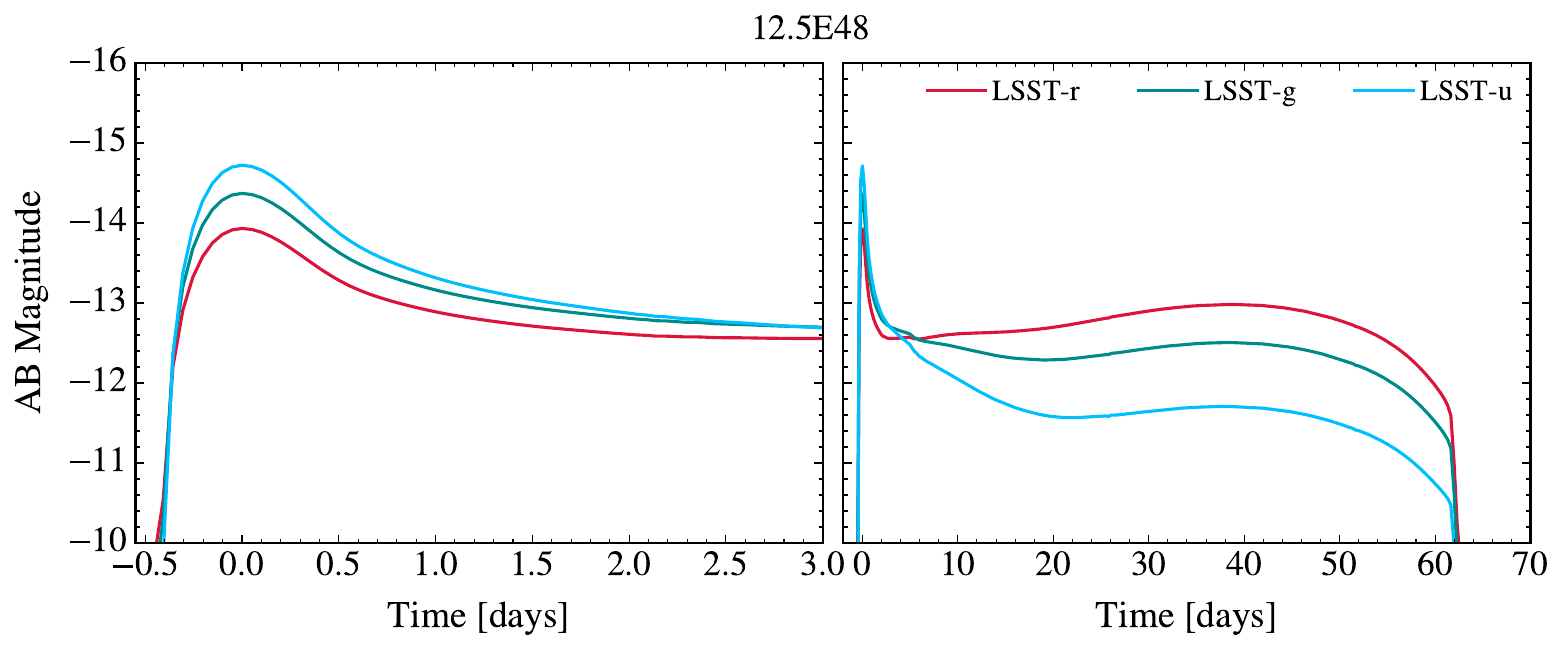}
\includegraphics[width=0.8\textwidth]{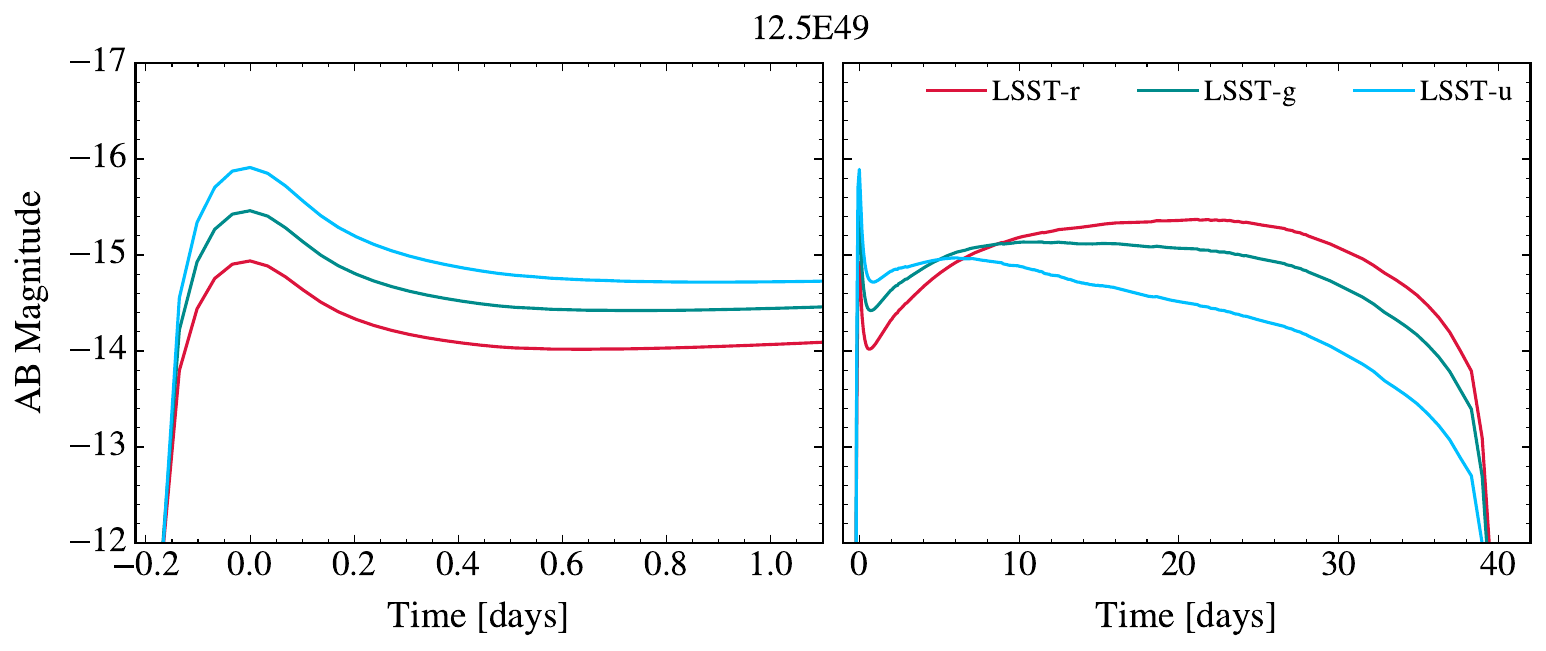}
\caption{Same as in Fig.~\ref{fig:broadband_12M_hrich} but showing the $\mzams = 12\msun$, $\mconv = 0.5\msun$ models with $\log(\extra / {\rm erg})$ = 48 (top row) and 49 (bottom row).}
\label{fig:broadband_12M_hpoor}
\end{figure*}

For each model, we use the instantaneous radial profiles that are output by \ath over the course of the simulation to calculate the radius of the photosphere, $R_{\rm ph}(t)$, and the bolometric luminosity just outside the photosphere, $L(t)$, with the photosphere defined to be the location where $\tau = 1$.  We then estimate $\teff$ using the Stefan-Boltzmann law 
$\teff = \big[L/({4\pi\sigma_{\rm sb}R_{\rm ph}^2})\big]^{1/4}$.  At each time, we assume the photosphere is a black body so that the luminosity at frequency $\nu$ is
\begin{equation}
L_\nu = {4\pi R_{\rm ph}^2\pi B_\nu(\teff)}
\end{equation}
with $B_\nu(T_{\rm eff})$ the Planck function.
Neglecting cosmological redshift, the flux density  at a distance $D$ in band $i$ is
\begin{equation}
f_i(D) = \frac{\int_{\nu_1}^{\nu_2}(L_\nu / 4\pi D^2) S_i(\nu) d\nu}{\int_{\nu_1}^{\nu_2}S_i(\nu) d\nu}
\end{equation}
where $S_i(\nu)$ is the response function for the photometric band.  Then the absolute magnitude in band $i$ (in the AB system) is 
\begin{equation}
    M_i = -2.5\log\bigg(\frac{f_i(D=10{\rm pc)}}{\rm 3631 Jy}\bigg).
\end{equation}
The broadband lightcurves for our $\mzams=12\msun$ RSG explosions are shown in Fig.~\ref{fig:broadband_12M_hrich}. Fig.~\ref{fig:broadband_12M_hpoor} shows the lightcurves for explosions of the H-poor star with $\mzams =12\msun$ and $\menv=0.5\msun$.

\bibliography{references}{}
\bibliographystyle{aasjournalv7}



\end{document}